\documentclass[onecolumn]{aa}  
\usepackage{hyperref}
\usepackage{graphicx}
\usepackage{txfonts}
\usepackage{amsmath}
\usepackage{color}
\usepackage{bm}
\usepackage{natbib}
\usepackage{listings}
\usepackage{algorithmic}
\usepackage[normalem]{ulem}
\usepackage[running]{lineno}

\makeatletter
\renewcommand*\aa@pageof{Page \thepage{} of \pageref*{LastPage}}

\makeatother

\def\be{\begin{equation}}
\def\ee{\end{equation}}

\newcommand{\oen}[1]{\textcolor{black}{#1}}

\usepackage[T1]{fontenc} 
\newcommand{\diff}{{\rm d}} 
\newcommand{\red}[1]{{\color{red}{#1}}} 

\makeatletter
\newcommand*{\rom}[1]{\expandafter\@slowromancap\romannumeral #1@}
\makeatother
\begin{document}

   \title{asevolution: a relativistic $N$-body implementation of the (a)symmetron} 

   \author{Øyvind Christiansen
          \inst{1}\fnmsep\thanks{Corresponding author's \email{oyvind.christiansen@astro.uio.no}},
          Farbod Hassani\inst{1}, Mona Jalilvand\inst{1,2,3} \and David F. Mota\inst{1}
          }

   \institute{Institute of Theoretical Astrophysics, University of Oslo,
             Sem Sælands vei 13, 0371 Oslo, Norway
         \and
             Department of Physics, McGill University, 3600 rue University, Montreal, QC H3A 2T8, Canada
             \and              
McGill Space Institute, McGill University, 3550 rue University, Montreal, QC H3A 2A7, Canada
             }


  \abstract
{We present asevolution, a cosmological $N$-body code developed based on gevolution, which consistently solves for the (a)symmetron scalar field and metric potentials within the weak-field approximation. In asevolution, the scalar field is dynamic and can form non-linear structures.
A cubic term is added in the symmetron potential to make the symmetry-broken vacuum expectation values different, which is motivated by observational tensions in the late-time universe. To study the effects of the scalar field dynamics, we also implement a constraint solver making use of the quasi-static approximation, and provide options for evaluating the background evolution, including using the full energy density averaged over the simulation box within the Friedmann equation. The asevolution code is validated by comparison with the Newtonian $N$-body code ISIS that makes use of the quasi-static approximation. There is found a small effect of including relativistic and weak-field corrections in our small test simulations; it is seen that for small masses, the field is dynamic 
and can not be accurately solved for using
the quasi-static approximation; 
and we observe
the formation of unstable domain walls and demonstrate a useful way to identify them within the code. A first consideration indicates that the domain walls are more unstable in the asymmetron scenario. }

    \titlerunning{asevolution, an $N$-body code solving the (a)symmetron}
    \authorrunning{Christiansen et. al}
   \maketitle
%
\section{  Introduction}

Withstanding decades of experimental verification, the relatively simple Lambda Cold Dark Matter ($\Lambda$CDM) model has successfully captured in its description most, if not all of the cosmological observations \citep{planck_planck_2020, Planck:2015fie, Pan-STARRS1:2017jku, BOSS:2016wmc}. However, 
the new generation of cosmological observations over the last decade indicates that the standard model of cosmology
fails to simultaneously fit well the late and early time cosmology, as exemplified in particular by the Hubble tension, a disagreement in the best fit for the Hubble parameter that has been claimed to reach a 5 sigma significance \citep{riess_comprehensive_2022}. Furthermore, there have been reported observations that seem to contradict the cosmological principle, a fundamental assumption in the $\Lambda$CDM model.
Recently, there has been a claim of a 5 sigma significance difference in the magnitude of the Cosmic Microwave Background (CMB) and the Large Scale Structure (LSS) dipole \citep{secrest_test_2021,secrest_challenge_2022}. We  expect a dipole to be present due to our peculiar velocity relative to the CMB rest frame, but we also expect the relative velocities of large structures to the CMB to be close to zero, since there should not be large scale coherent bulk flows \citep{planck_bulk}; If the observation of \cite{secrest_test_2021} indicating otherwise is not a result of systematic errors, it could herald the need for new physics. 

We set out here to suggest and explore a candidate cosmological model that has already been studied somewhat recently, using spherical collapse \citep{perivolaropoulos_gravitational_2022} -- the asymmetron, in which we add a cubic term to the potential of the more well-known symmetron dark energy model of \cite{hinterbichler_symmetron_2010}, so that the field may find different Vacuum Expectation Values (VEV) after symmetry breaking.
The model has a natural screening mechanism that couples the shape of the potential to the matter energy density, making the VEV vanish in high density environments where there are otherwise strict constraints from highly accurate experiments in the solar system \citep{Will:2001mx,Bertotti:2003rm, Esposito-Farese:2004azw,1987MNRAS.227....1K}. Choosing parameters that maintain the parity symmetry of the symmetron model for early redshifts (z > 100), we may avoid unwanted departures from the verified predictions of $\Lambda$CDM in observables such as the CMB \citep{ivanov_constraining_2020,zhang_ede_2021}. The asymmetron causes a partitioning of the Universe into separate regions, each of which is characterised by a stable potential minimum and bordered by domain walls. The scale of the resulting domains may be large if the Spontaneous Symmetry Breaking (SSB) happens sufficiently late \citep{llinares_domain_2014}. These domains may induce real clustering asymmetries in the large-scale structure through their different minima's influence on the fifth force strength, and observational asymmetries through secondary effects. We aim to study these effects in the future. The calibration parameters of standard candles, such as cepheids or the tip of the red giant branch, located in different domains will become environmentally dependent and cause systematic errors to appear in inferences not taking them into account, possibly causing the Hubble tension \citep{desmond_local_2019,desmond_screened_2020,perivolaropoulos_hubble_2021,alestas_hints_2021}.

To study the asymmetron scenario, we develop asevolution, a relativistic $N$-body code that consistently evolves the particles by solving the metric degrees of freedom and the asymmetron scalar field. This approach to exploring the complex phenomena
in cosmology that is not analytically accessible has a rich past. During past years various $N$-body simulations have been developed to study the non-linear structure formation in different theories of gravity \citep{Baldi:2008ay,Barreira:2013eea,llinares_isis_2014,Adamek:2017uiq, Winther:2017jof,LiBook,hassani_k-evolution_2019,Hassani:2020rxd, brax_systematic_2012}.
The $N$-body code RAMSES \citep{teyssier_cosmological_2002,teyssier_ramses_2010} specifically, is an Adaptive Mesh Resolution (AMR) code that has been widely used for studying both cosmological and galactic scales. It has also been applied to modified gravity scenarios \citep{li_ecosmog_2012,brax_systematic_2012,llinares_isis_2014}, including the symmetron. Furthermore, the implementation of the symmetron has been extended in \cite{llinares_releasing_2013} and \cite{llinares_cosmological_2014} to include its dynamic part and they have found small effects on the local power spectrum, and a negligible effect on the global power spectrum. However, in the previously mentioned references only a portion of the parameter space is considered, restricted in part by they only keeping leading orders of the conformal factor. The smallness of the conformal factor is motivated by an argument made from local solar system experiments in \cite{hinterbichler_symmetron_2010}, which we aim to consider more carefully. They were also evolving the fields using Newtonian gravity. This work is an extension of \cite{llinares_releasing_2013,llinares_cosmological_2014} and additionally sets the path towards identifying other possible observables on which the (a)symmetron might have a larger impact. The cubic term in the potential has not been implemented before either, and it can in principle help in evading the constraints made in \cite{hinterbichler_symmetron_2010,nagata_wmap_2004} by allowing two minima for the asymmetron, of which only one needs to satisfy local experiments. We also consider whether the non-linear and dynamic structure of the asymmetron might have a significant level of back-reaction, which would make it necessary to evolve the background by averaging over the full energy densities. The gevolution code has been used for studying cosmological back-reaction for the $\Lambda$CDM scenario in the past, finding no important effects in the Poisson gauge \citep{adamek_safely_2019}.

Our aim of extending the rigour of previous implementations with asevolution can be viewed in the context of making optimal usage of the discriminative power of recent and upcoming cosmological surveys such as Euclid \citep{racca_euclid_2016}, Square-Kilometre-Array (SKA) \citep{weltman_fundamental_2020} and the Dark Energy Spectroscopic Instrument \citep{desi_2016}. To do so, it is necessary to have equally high-precision modeling of the relevant physics. Therefore, it seems relevant to include first-order corrections in the metric potentials, to keep all orders in the velocity, and to consider the full system of coupled differential equations for the asymmetron field. For this, we choose to modify the recently released Particle-Mesh (PM) $N$-body code gevolution \citep{adamek_general_2016,adamek_gevolution_2016}, that successfully and efficiently implements a fully relativistic and linear order in metric potentials $N$-body code for $\Lambda$CDM and dark fluid cosmologies. Furthermore, it has on-the-fly lightcone production that will be useful in forecasting and producing mock catalogues for comparison with observational data. Notably, there has also been developed a relativistic implementation of RAMSES, GRAMSES \citep{barrera-hinojosa_gramses_2020} that makes use of Adaptive Mesh Refinements (AMR) that can improve on computational costs for resolving small scales. For implementing the asymmetron, we will follow the general modification scheme of $k$-evolution \citep{hassani_k-evolution_2019, hassani_parametrising_2020} that similarly introduces an extra scalar field, which in their case is minimally coupled to gravity. The gevolution code has also been applied to $f(R)$ gravity in the past \citep{reverberi_frevolution_2019}.

\subsection{Conventions}\label{SS:units}
In the following, we will set the speed of light equal to 1
and use the metric signature $(-+++)$. $A$ is the conformal factor and has implicit dependence on the scalar field $\phi$. We introduce the notation $h^{-1}$cMpc for comoving megaparsec, in order to reduce ambiguity around whether or not a quantity is comoving or physical. All quantities with $\sim$ on top are defined in the Jordan frame, such as the Jordan frame matter energy density $\tilde\rho_m$; In the Einstein frame, we write $\rho_m$.

\section{Theory}\label{S:theory}
We define the asymmetron action in the Einstein frame similarly to how the symmetron is defined in \cite{hinterbichler_symmetron_2010}, namely as the Einstein-Hilbert action plus the Klein-Gordon Lagrangian for the asymmetron field
\begin{linenomath}
\begin{equation}\label{eq:action}
S  = \int \diff^{4} x \sqrt{-g} \, \Bigg[\frac{1}{2} M_{\text{pl}}^2 R-\frac{1}{2} \nabla_{\mu} \phi \nabla^{\mu} \phi-V(\phi)\Bigg]+ S_{m}\left(\psi_{i}, \tilde{g}_{\mu \nu}\right),
\end{equation}
\end{linenomath}
where $R$ is the Ricci scalar and $g$ is the determinant of the metric; $X=-\frac{1}{2} \nabla_{\mu} \phi \nabla^{\mu} \phi$ is the kinetic energy of the asymmetron field $\phi$ and $V(\phi)$ is its potential energy; $M_{\text{pl}}=1/\sqrt{8\pi G_N}$ is the Planck mass while $G_N$ is the Newtonian gravitational constant; and $S_m$ is the Jordan frame matter action, expressed as a function of the matter fields $\psi_i$ and the Jordan frame metric $\tilde g_{\mu\nu}$.
The action is stated in the Einstein frame, which means that the scalar field is minimally coupled to gravity and that the matter action will contain interactions with the asymmetron field; In the Jordan frame the matter action is that of the standard model. We can obtain the Einstein frame matter action by applying the conformal transformation
\begin{linenomath}
\begin{align}\label{eq:conformaltransform}
\tilde g_{\mu\nu}=A^2(\phi)g_{\mu\nu},
\end{align}
\end{linenomath}
where $g_{\mu\nu}$ is the metric tensor in the Einstein frame and the conformal factor is defined as
\begin{linenomath}
\begin{align}\label{eq:conformalfactor}
A(\phi) \equiv 1 + \frac{1}{2}\left(\frac{\phi}{M}\right)^2 \equiv 1 +\Delta A,
\end{align}
\end{linenomath}
in which $M$ is the conformal coupling scale and is a free parameter of the theory.
Performing the variation of the Lagrangian density $\mathcal{L}_m$ with respect to the scalar field, one can find the term from the matter action that will source the equations of motion of the scalar field
\begin{linenomath}
\begin{align}
\frac{\delta \mathcal{L}_m}{\delta \phi} = \frac{\delta \mathcal{L}_m}{\delta \tilde g^{\mu\nu}}\frac{\delta\tilde  g^{\mu\nu}}{\delta \phi} = -\frac{\sqrt{-\tilde g}}{2}\tilde T A^2 \partial_\phi \left(A^{-2}\right) = \sqrt{-g} \,A^3 \tilde T A_\phi,
\end{align}
\end{linenomath}
where $\tilde T$ is the trace of the stress-energy tensor in the Jordan frame, defined as $\tilde g^{\mu\nu}\tilde T_{\mu \nu}\equiv -\tilde g^{\mu\nu}\, 2/\sqrt{-\tilde g}\, \delta \mathcal{L}_m /\delta \tilde g^{\mu \nu}$, and $\mathcal{L}_m$ is the Lagrangian density of matter. Expressing this with the Einstein frame Stress Energy (SE) tensor $T$ instead, using  the relation $A^4 \tilde T = T$, we have $\sqrt{-g}T A_\phi/A$. We can add a term to the potential to get an effective potential that will generate the correct equations of motion\footnote{For most parameter choices we can safely use $\ln A\sim A$ as the equations of motion would differ by a factor of A which for small $\Delta A$ is higher order. We are however only doing this approximation for the parameter mapping.}
\begin{linenomath}
\begin{equation}\label{eq:potential}
    V_{\text{eff}}  \equiv V - \ln A(\phi) T_m = V_0 - \frac{1}{2}\mu^2\phi^2 - \frac{1}{3}\kappa \phi^3 + \frac{1}{4}\lambda \phi^4 - \ln A(\phi) T_m,
\end{equation}
\end{linenomath}
for which we have defined the mass $\mu$, the cubic coupling\footnote{This choice of sign for the $\kappa$ term allows us to use both $\beta_+>\beta_-$ (will be defined later) and $\kappa>0$ conventions.} $\kappa$, the $\phi^4$ coupling $\lambda$ and the constant $V_0$. Note that $V_0$ will have the effect of the cosmological constant, whereas $\kappa$ is the modification performed in \cite{perivolaropoulos_gravitational_2022} that when put to zero gives back the symmetron model. The SE tensor trace $T$ in the effective potential should be treated as a constant when performing the variation with respect to $\phi$. Expressed with the Jordan frame matter stress energy tensor we have instead $-\ln A(\phi) T_m \sim - A^4 \tilde T_m/4$, which generates the same equations of motion.
\subsection{Phenomenological parameters}
\begin{figure}[!ht]
    \centering
    \includegraphics[width=0.5\linewidth]{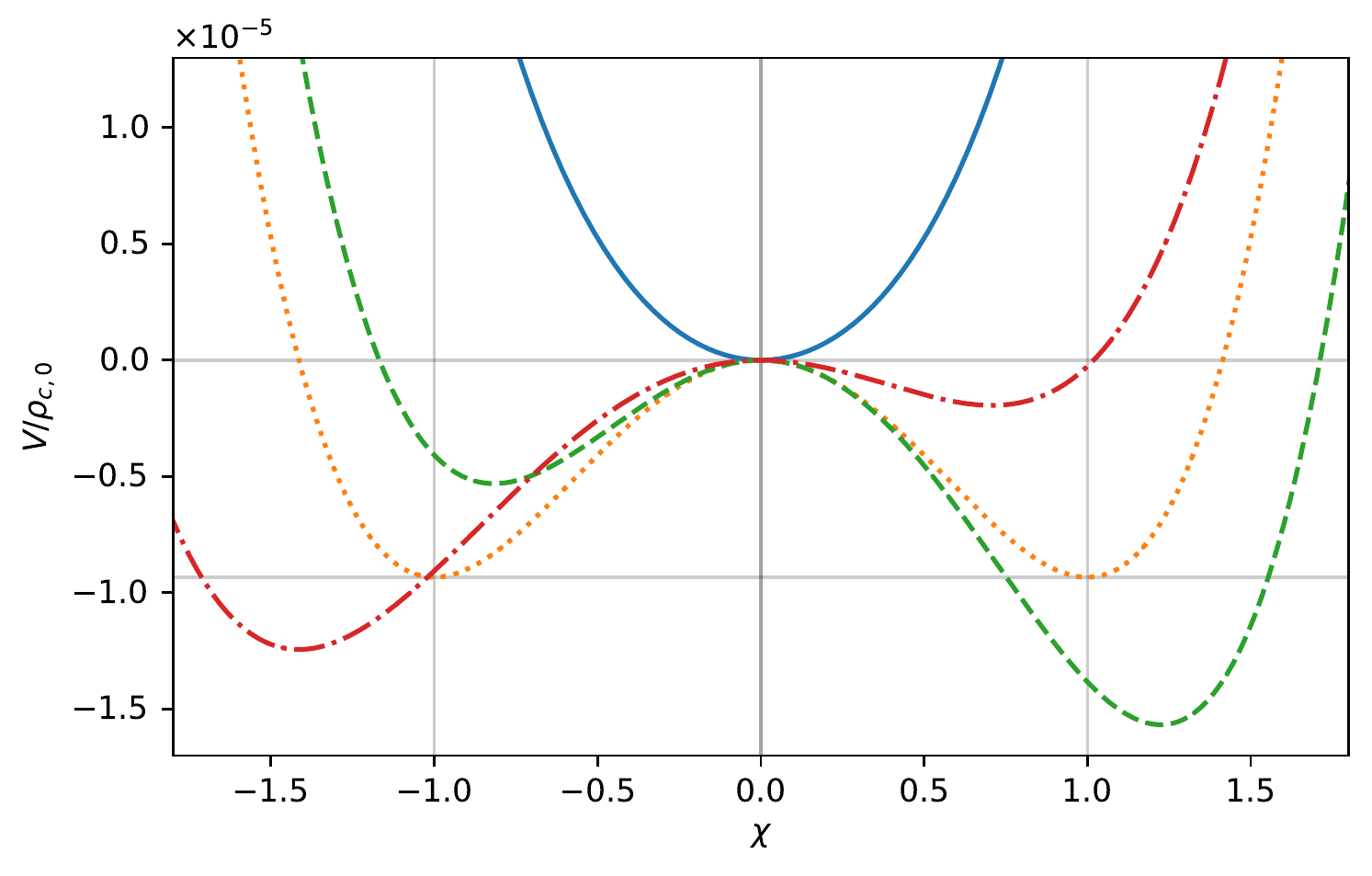}
    \caption{Possible shapes of potentials and minima for different choices of model parameters. The potentials are shown here as a function of $\chi = \phi/v$. Any asymmetry across the y-axis is given by the cubic term. All curves correspond to parameters $(L_*,a_*) = (1 \text{ Mpc/h},0.33)$. The blue solid line has positive sign for the mass term and $\beta_+=\beta_-=1$. The rest are plotted for $\rho_m=0$ and have parameters $(\beta_+,\beta_-)$ equal to $(1,1)$ for orange dotted line, $(1.2,0.8)$ for green dashed line, and $(1,0.5)$ for red dashdotted line which also has $\kappa<0$. The y-axis is the energy density of the potential relative to the critical density at redshift $z=0$.}
    \label{fig:achipotential_qualitative}
\end{figure}

As described in \cite{brax_systematic_2012,davis_structure_2012}, we can map phenomenologically relevant quantities to the free parameters of the potential using the following scheme:  $ (\xi_*,a_*,\beta_+,\beta_-) \rightarrow (\mu,M,\kappa,\lambda)$, where $\xi_*$ represent the Compton wavelength as a fraction of the Hubble length, $a_*$ is the scale factor at symmetry breaking, $\beta_+$ is the fifth
 force strength relative to the Newtonian gravitational force and $\beta_-$ is the second of the two fifth force strengths. For the symmetron, where $\beta_*\equiv \beta_+=\beta_-$, we find (see appendix \ref{A:mapping})\footnote{Equations \eqref{eq:map1} and \eqref{eq:map2} both are true for the asymmetron case too, whereas \eqref{eq:map3} needs to be generalised. This is done in  appendix \ref{A:mapping}.}
 \begin{linenomath}
\begin{align}\label{eq:map1}
    L_{C} &= \frac{1}{\sqrt{2}\mu} \equiv \frac{\xi_*}{H_0} \approx \xi_* \cdot 2998 \text{ Mpc/h}, \\\label{eq:map2}
    M &= \xi_*\sqrt{6\Omega_{m,0} /a^{3}_*},\\\label{eq:map3}
    \lambda &= \frac{H_0^2}{72\Omega_m^2}\left(\frac{a_*^3}{\beta_*\xi_*^3}\right)^2,
\end{align}
\end{linenomath}
where we have approximated the trace of the matter stress-energy tensor with the matter background energy density $\rho_m=\rho_{c,0}\Omega_m a^{-3}$ and defined the Compton wavelength $L_C$. We will show how to derive the mapping for $M$ here and leave the rest to appendix \ref{A:mapping}. We define the symmetry breaking to happen when the shape of the effective potential changes so that the only stable minima are at non-zero field values. Differentiating the potential with respect to the field, expanding in small $\Delta A$, we find 
\begin{linenomath}
\begin{align} \label{eq:div-potential}
    \partial_\phi V_{\rm{eff}} = \lambda\phi^3 + \phi \mu^2 \left( 
    \frac{\rho_m}{M^2\mu^2}-1
    \right) - \kappa \phi^2. 
\end{align}
\end{linenomath}
Evaluating the derivative at $\phi=0$, we immediately find that it has a trivial zero value. To determine the stability of the potential there,  we calculate 
\begin{linenomath}
\begin{align}\label{eq:SSBpotential}
    \left. \partial^2_\phi V_{\rm{eff}}\right|_{\phi=0} \propto \frac{\rho_m}{M^2\mu^2} - 1,
\end{align}
\end{linenomath}
from which we realise that the effective potential becomes unstable when $\rho_m < M^2\mu^2$, which is the moment we define as the spontaneous symmetry breaking, where the field will start moving towards the two new minima. Note however that the redshift at which symmetry breaking occurs, is independent of the cubic term, although the field values at the minima are dependent on it.
We can find the vacuum expectation values by solving for $\phi$ where  \eqref{eq:div-potential} is zero, which gives 
\begin{linenomath}
\begin{align} \label{eq:potentialZeros}
    \phi_0 = 0 \quad \vee \quad \phi_\pm = \frac{\kappa \pm \sqrt{\kappa^2 - 4 \lambda \mu^2\left(
    \frac{\rho_m}{M^2\mu^2}-1
    \right)}}{2\lambda}.
\end{align}
\end{linenomath}
For the symmetron case we have
\begin{linenomath}
\begin{align}\label{eq:expectation_sym}
    \phi|_{\kappa=0}
    = v \sqrt{1-\left(a_*/a\right)^3},
\end{align}
\end{linenomath}
and for the vacuum case at background level
\begin{linenomath}
\begin{align}\label{eq:vev}
    \phi_\pm|_{\rho_m=\kappa=0} \equiv \pm v = \pm \frac{\mu}{\sqrt{\lambda}}, 
\end{align}
\end{linenomath}
where $v$ is the symmetron vacuum expectation value.
For $\kappa=0$, we find that the non-zero minima are complex valued before symmetry breaking (if we allow non-zero $\rho_m$), so the field which is $\phi\in \mathbb{R}$ can not take on these values. The addition of a cubic term will allow them to become real at an earlier time. However, the centre minimum remains stable until the symmetry breaking time defined in equation \eqref{eq:potentialZeros}.
More generally, we define the asymmetron vacuum expectation values 
\begin{linenomath}
\begin{align}\label{eq:avev}
    v_\pm \equiv \left|\frac{\kappa \pm \sqrt{\kappa^2+4\lambda\mu^2}}{2\lambda}\right|.
\end{align}
\end{linenomath}
The behaviour of some different choices of potentials is shown in figure \ref{fig:achipotential_qualitative}.
Next, we give a quick way to understand the formula for the fifth force. We realise that in the Jordan frame the fifth force is part of the gravitational force, and it is the Bardeen potential $\Psi$ of the time-component of the metric in the Newtonian gauge that sources the Newtonian limit gravitational force. If we assume that $\left(\phi/M\right)^2\ll 1$, that the time derivative of the scale factor $\mathcal{H}\equiv \dot a \ll 1$, and that the Einstein frame fifth force arises from the conformal transformation of the Jordan frame Newtonian gravitational potential $\Psi$, then the acceleration $\ddot x$ of a test particle is
\begin{linenomath}
\begin{align}
    \ddot x = -\partial_r \tilde\Psi  = -\partial_r\left[ \Psi + \Delta A\right] = -\partial_r\Psi
    - \frac{\phi}{M^2}\partial_r\phi,
\end{align}
\end{linenomath}
where we have used the transformation of the Bardeen potential that we will state later in equation \eqref{eq:metrictransform1}. The last term, $-\phi\partial_r\phi /M^2$, is the correct formula for the quasi-static, Newtonian limit fifth force, according to among other \cite{davis_structure_2012}. We then restate it as
\begin{linenomath}
\begin{align} \label{eq:fifthforce}
    F_{5,\pm} = -\frac{\beta_{{\pm}}}{M_{\text{pl}}}\left(\frac{\phi}{v_{{\pm}}} \right)\partial_r \phi,
\end{align}
\end{linenomath}
where we have used $\beta_{{\pm}}$, which is defined as
\begin{linenomath}
\begin{align}\label{eq:betadef}
    \beta_{\pm} \equiv \frac{v_\pm M_{\text{pl}}}{M^2}.
\end{align}
\end{linenomath}
For the symmetron case when $\kappa=0$, we define $\beta_*\equiv \beta_+=\beta_-$, which has been shown in the case of spherical symmetry by \cite{hinterbichler_symmetron_2010} to give the relative force strength  $F_5/F_N = 2\beta_*^2$.
The definition \eqref{eq:betadef} does not change as we introduce the cubic term in the potential, as the fifth force only depends on the conformal coupling explicitly. We see that there is an implicit dependence on the other parameters through the vacuum expectation value. If we define it to be the one of the asymmetron in equation \eqref{eq:avev}, we find that $\beta_{{\pm}}$ indeed can have two values when $\kappa\neq 0$, so that there will be different fifth force strengths in the two minima after SSB. We define these hereafter $v_\pm$, $F_\pm$ and $\beta_\pm$, that for the symmetron case become degenerate. Since we defined the VEV like in equation \eqref{eq:avev}, for positive $\kappa$, we will always have $F_+>F_-$.

\section{Implementation}\label{S:implementation}

In this section we will review the different choices of schemes made for the implementation into gevolution. We discuss the background, the Einstein field equations, the equations of motion of the asymmetron, the alternative constraint solver for the asymmetron, the stress energy tensor of the field and the effect on the geodesic equation of the dark matter particles.
\subsection{Metric}
We will use the Poisson gauge, where we can state the space time interval as
\begin{linenomath}
\begin{align}\label{eq:poissongauge}
    \diff s^2 = a(\tau)^2\Big[ -e^{2\Psi (\tau, \vec x)}\diff \tau^2 
    - 2 B_i(\tau, \vec x) \diff x^i \diff \tau + (e^{-2\Phi
    (\tau, \vec x)}\delta_{ij} + h_{ij}(\tau, \vec x))\diff x^i \diff x^j \Big],
\end{align}
\end{linenomath}
where we use conformal time $\tau = \int \diff t/a$ instead of cosmic time so that the scale factor factors out. The Poisson gauge is fixed using the following conditions on the metric perturbations
\begin{linenomath}
\begin{align}\label{eq:gaugeconditions}
    h = \delta^{ij}h_{ij} = 0,\quad \partial_i h^{ij} = 0,\quad \partial_i B^i = 0,
\end{align}
\end{linenomath}
which eliminates $1+3+1=5$ degrees of freedom\footnote{4 additional degrees of freedom can be eliminated using the Bianchi identities, which gives us the two physical massless tensor degrees of freedom.} and leaves us with 6. 
We will treat the metric perturbations perturbatively according to the linearisation scheme of \cite{adamek_gevolution_2016}, where the smallness parameter $\epsilon$ is 
\begin{linenomath}
\begin{align}\label{eq:epsilon}
    \epsilon\sim \Phi\sim \Psi\sim B_i \sim h_{ij}.
\end{align}
\end{linenomath}
For $\Psi$ and $\Phi$, spatial derivatives lower the order of the term by $\epsilon^{1/2}$ which can be understood by considering the source of the relativistic Poisson equation which is of order 1, due to non-linear structure formation in the late time. Time derivatives do not affect the counting, and furthermore the velocity is treated non-perturbatively, i.e. $v\sim 1$.

\subsection{Einstein and Jordan frames}

We have found it simplest to implement the (a)symmetron model in the Einstein frame, where the scalar field is added to the matter sector. In this frame, the (a)symmetron field has a Klein-Gordon free-field action which is sourced by the effective potential, equation \eqref{eq:potential}, containing the interaction with the standard model matter fields. In this frame, gravitation is governed by the standard Einstein-Hilbert action. However, as we will discuss in section \ref{SS:fifthforce}, we are evolving the particles in the Jordan frame, since in this frame the interaction with the standard model fields is mediated through gravity, and we thus find the particles' complete dynamics by solving the geodesic equation. 
We can transform the metric potentials to the Jordan frame ones by
\begin{linenomath}
\begin{align}\label{eq:metrictransform1}
    \tilde \Psi \,&= \Psi + \Delta A
    ,\\
    \tilde \Phi\, &= \Phi- \Delta A
    ,\\
    \tilde h_{ij} &=  h_{ij},\quad
    \tilde B_i\, =  B_i,\label{eq:metrictransform-1}
\end{align}
\end{linenomath}
which one can easily derive through equation \eqref{eq:conformaltransform} and expanding in $\Delta A\equiv A-1$ assuming $\Phi\Delta A,\left(\Delta A\right)^2 \ll \Phi$. If the Einstein frame linearisation scheme of equation \eqref{eq:epsilon} is also to hold for the Jordan frame, we need $\Delta A \lesssim \Phi$. We will use $\partial_i A \lesssim \partial_i \Phi$ since the fifth force can become up to the same order as the gravitational force. It is also assumed that $\dot A\lesssim \dot\Phi$, although the oscillatory nature of the field might make it interesting to consider relaxing this condition in the future. If $\Phi\sim 10^{-5}$ at cosmological scales, and as a result $\Delta A \lesssim 10^{-5}$, then assuming that the field is in its vacuum \eqref{eq:vev}, using the parameter mapping presented in appendix \ref{A:mapping}, we have
\begin{linenomath}
\begin{align}\label{eq:symmetronGeodesicConstraint}
    \Delta A \sim \frac{1}{2}\frac{\mu^2}{\lambda M^2} = 3 \Omega_m \left( \frac{\beta_*^2\xi_*^2}{a_*^3} \right) \sim \frac{\beta_*^2\xi_*^2}{a_*^3}\lesssim 10^{-5},
\end{align}
\end{linenomath}
which constrains the available parameter space of our implementation. For $\beta_*\sim 1$ and $a_*\sim 0.1$, we would require $\xi_*\lesssim 10^{-4}$ or the Compton wavelength $L_C \lesssim 0.3$ Mpc/h. Using a later symmetry breaking $a_*\sim 0.5$, would allow $L_C\lesssim 3$ Mpc/h. However, if we allow smaller fifth forces, or $\beta$'s, we can consistently evolve the particles for larger Compton wavelengths. Since the requirement to keep the Jordan frame potentials small only applies to the implementation of the fifth force, our implementation will still allow an exploration of other effects such as the ones on the background evolution for an extended parameter space.

The Jordan frame potentials are used for the evolution of the particles, which are immediately brought back to the Einstein frame so that the matter power spectrum can be compared with the one from the ISIS $N$-body code \citep{llinares_isis_2014} that is also using the Einstein frame. While in the past there has been some confusion about the equivalence of the two frames, it has later been demonstrated that their effect on observables is the same  \citep{deruelle_conformal_2011,francfort_cosmological_2019,francfort_observables_2022}.
\begin{figure}[!ht]
    \centering
    \includegraphics[width=0.6\linewidth]{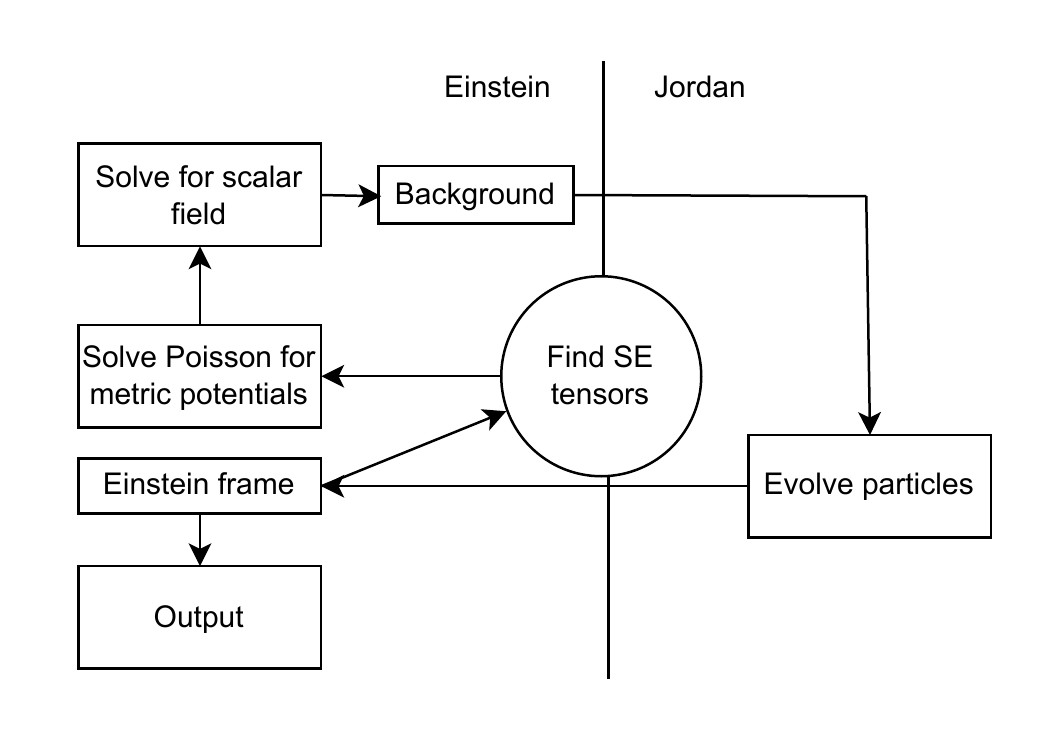}
    \caption{Flowchart diagram displaying some of the main modules of asevolution and which frame they are performed in.}
    \label{fig:FlowChart}
\end{figure}
In figure \ref{fig:FlowChart} we show a flow chart indicating the general structure of asevolution, and what is happening in each frame. One should be careful to use the relevant frame metric potentials. The stress-energy tensor is written in the middle, since the total SE tensor is expressed as the sum of the scalar field part and the Jordan frame standard matter SE tensor transformed to the Einstein frame (see section \ref{SS:SEtensor}). The output in asevolution is made in the Einstein frame, which is important to specify as we only expect observables to be independent of frame, and not for example the matter power spectrum. To get the dimensionless  matter power spectrum, $\mathcal{P}_{m}$, in the Einstein frame, we can make use of the transformation presented in \cite{francfort_cosmological_2019}
\begin{linenomath}
\begin{align}\label{eq:PkTransform}
    \mathcal{P}_{m} = \tilde{\mathcal{P}}_{m} - 8 \delta_{m} \frac{\Delta A}{A} - 16 \left(\frac{\Delta A}{A}\right)^2,
\end{align}
\end{linenomath}
where everything is evaluated in $k$-space and $\delta_m$ is the matter over density.

\subsection{Background}
For the background, we have implemented a linearisation scheme similarly to what was done in \cite{brax_systematic_2012}, however, since we are here interested in studying inherently non-linear phenomenology, we place ourselves agnostically regarding the effect of non-linearities on the background evolution and do not want to constrain ourselves needlessly to any subset of the parameter space. We therefore choose to also use the average of the energy density, which is already evaluated within the code as a source for the Poisson equation, and integrate the Friedmann equation directly as
\begin{linenomath}
\begin{align}\label{eq:friedmann}
    \dot a \equiv \mathcal{H} = H_0\sqrt{\frac{8\pi G_N}{3}a^2\rho_{\text{tot}}},
\end{align}
\end{linenomath}
with $a$ being the scale factor, $\mathcal{H}$ the conformal Hubble parameter, $H_0$ the Hubble parameter at $z=0$ and $\rho$ the energy density. Since we are in the Einstein frame, the full effect of the additional degree of freedom on the background evolution is captured by the unmodified Friedmann equation \ref{eq:friedmann}, with the additional energy density contribution of the scalar field. In section \ref{SS:backgroundresults}, we will compare the background evolution found from equation \ref{eq:friedmann}, computed from two different sources: 1) when inputted the box-averaged energy densities and 2) using the analytic background value \eqref{eq:expectation_sym} inserted into the non-effective potential $V$, equation \eqref{eq:potential}. In the latter case, we use the standard scale factor power laws for the standard model fields' energy density parameters
\begin{linenomath}
\begin{align}\label{eq:symbackground}
    \rho^{(S)}_{\text{tot}} = V\left(\bar\phi\right) + \frac{1}{2}\left(\frac{\bar\phi}{M}\right)^2 \Omega_{m,0}/a^3 + \Omega_{m,0}/a^3 + \Omega_{r,0}/a^4 + \Omega_{\Lambda,0} ,
\end{align}
\end{linenomath}
where we have used units $\rho_{c,0}=1$ and $\Omega_m,\Omega_r,\Omega_\Lambda$ are energy density parameters for matter, radiation and vacuum energy respectively. Since $\Omega_\Lambda$ is a free parameter of $\Lambda$CDM, we will adjust it for each individual run to maintain zero curvature $\Omega_k=0$ so that $\sum_i \Omega_i$=1. The contribution of the symmetron is shown in equation \eqref{eq:backgroundsymcontribution}, and as we comment there, we do not implement the asymmetron at background level since there is an ambiguity of which minimum to choose. Apart from this, the generalisation is trivial. The prefactor in front of the interaction term $1/2$ might be surprising, since we will show that $\rho=A^4\tilde \rho$; From the expansion of $A$, we would expect a prefactor $2$. The prefactor $1/2$ may be viewed as owing to the transformation of the scale factor, since at the background level, instead of absorbing $\Delta A$ into the perturbations, we may add it to the scale factor $a = A \tilde a$. We will discuss this more precisely in subsection \ref{A:energytransfercheck}.
Finally, we add an option to evolve the background dynamically according to the averaged equations of motion \eqref{eq:tmp1}.  The initial conditions are first set from taking averages of the initial conditions of the simulation, and then we integrate the velocity of the field with a fourth order Runge-Kutta method. The scalar field is evolved with the Euler-Cromer method. Equation \eqref{eq:symbackground} then has an additional term $\dot{\bar\phi}^2/2$, the importance of which is elaborated on in \cite{brax_unified_2012}. 

\subsection{Equations of motion}

We find the equations of motion for the scalar degree of freedom applying the Euler-Lagrange equation in curved spacetime on the Lagrangian stated in equation \eqref{eq:action}. We will first find the unperturbed\footnote{Meaning unperturbed in the metric.} FLRW equations of motion and then extend them to the perturbed ones, linearly in the metric potentials. The Euler-Lagrange equation is (with $\nabla$ being the covariant derivative defined with the Christoffel connection)
\begin{linenomath}
\begin{align}\label{eq:EL}
    \frac{\delta \mathcal{L}}{\delta \phi} - \nabla_\mu \frac{\delta \mathcal{L}}{\delta \left(\partial_\mu \phi\right)} = 0,
\end{align}
\end{linenomath}
which evaluates to $\square \phi  = \partial_\phi V_{\rm{eff}}$, and in the unperturbed FLRW case and cosmic time coordinates  we find 
\begin{linenomath}
\begin{align} \label{eq:tmp1}
    \ddot \phi + 3 H \dot \phi - \frac{\nabla^2\phi}{a^2} = -\lambda \phi^3 + \kappa \phi^2 -\left( \rho_{m}/M^2 - \mu^2\right)\phi,
\end{align}
\end{linenomath}
where we have differentiated the effective potential, equation \eqref{eq:potential}, on right hand side and $H$ is the Hubble parameter. We have also done an expansion in $\left(v/M\right)^2$. We can do the same trick as is done in \cite{llinares_releasing_2013} and define the velocity\footnote{We are defining $\chi'$ to mean derivative with respect to conformal time.} of the field, $q$, and the dimensionless scalar field, $\chi$,
\begin{linenomath}
\begin{align}\label{eq:ELeq2wdef}
    q \equiv a^3 \dot \chi = a^2  \chi ',\quad \chi \equiv \frac{\phi}{v},
\end{align}
\end{linenomath}
which when inserted into \eqref{eq:tmp1} eliminates the Hubble factor and yields
\begin{linenomath}
\begin{align} \label{eq:ELeq1}
    \dot q = a \nabla^2\chi -a^3 \mu^2\left\{
    \chi^3 - \bar \kappa \chi^2 + \left(
    \eta  - 1
    \right) \chi
    \right\},
\end{align}
\end{linenomath}
where we have used the definition \eqref{eq:vev} and defined
\begin{linenomath}
\begin{align}
    \bar\kappa \equiv \frac{\kappa}{\mu\sqrt{\lambda}},\quad \eta(x) \equiv \frac{\rho_m(x)}{M^2\mu^2}.
\end{align}
\end{linenomath}
The equations \eqref{eq:ELeq2wdef} and \eqref{eq:ELeq1} define the system of equations that we solve to evolve the scalar field. To extend this result to being linear in the metric potential, we need to take into account that the Christoffel symbols of the perturbed spacetime, that enter into the covariant derivative of equation \eqref{eq:EL}. The perturbations are defined in the Poisson gauge stated in equation \eqref{eq:poissongauge}, and we will delay imposing the gauge conditions \eqref{eq:gaugeconditions} for now. We can find the inverse metric\footnote{We are raising and lowering indices of the perturbations with the Kronecker delta.}
\begin{linenomath}
\begin{align}
     g^{\mu\nu} = 
 a^{-2} \begin{pmatrix}
    \quad-e^{-2\Psi} & -B_i \\
    -B_i & \delta_{ij}e^{2\Phi} -h_{ij}
    \end{pmatrix},
\end{align}
\end{linenomath}
where we have used the determinant
\begin{linenomath}
\begin{align}
    g = a^8\left(-1-2\Psi+6\Phi-h\right).
\end{align}
\end{linenomath}
In the end, we find the Christoffel symbols\footnote{Note that these are with respect to the conformal time coordinates.}.
\begin{linenomath}
\begin{align}
    \Gamma^\alpha_{0\alpha} &= \frac{1}{2}g^{\lambda\alpha}g_{\lambda\alpha,0} = \partial_0 \left( \Psi-3\Phi+\frac{1}{2}h\right)+4\mathcal{H} , \\
    \Gamma^\alpha_{i\alpha} &=  \frac{1}{2}g^{\lambda\alpha}g_{\lambda\alpha,i} =\partial_i \left( \Psi-3\Phi+\frac{1}{2}h\right).
\end{align}
\end{linenomath}
We see that these reduce to the FLRW ones when the perturbations are zero. Left hand side of equation \eqref{eq:tmp1} then evaluates to 
\begin{linenomath}
\begin{align}
    \Box \phi = \nabla_\mu g^{\mu\nu} \partial_\nu \phi =  \left(\partial_\mu g^{\mu\nu}\partial_\nu +g^{\mu\nu}\Gamma^\alpha_{\mu\alpha}\partial_\nu\right) \phi ,
\end{align}  
\end{linenomath}
which gives us to the first order in the potentials
\begin{linenomath}
\begin{align}
    a^2\Box\phi = \left(-\left[ 1-2\Psi \right]\left[\partial^2_0 -2\mathcal{H}\partial_0 \right]+\left[1+2\Phi\right]\nabla  - 2 B^i\left[ \partial_i\partial_0 - 2\mathcal{H}\partial_i \right]- h^{ij}\partial_i\partial_j \right)\phi+ \left( -\left[ \Gamma^\alpha_{0\alpha} - 8\mathcal{H}\Psi  \right] \partial_0 - 4\mathcal{H}B^i\partial_i + \delta^{ij}\Gamma^\alpha_{i\alpha}\partial_j\right)\phi.
\end{align}
\end{linenomath}
Dividing both sides on the symmetron VEV $v$ and writing this in terms of the field $q$, we see that the Hubble factor disappears
\begin{linenomath}
\begin{align}
    a^4\Box\chi =  -\left[1-2\Psi\right]q' - 2 B^i \partial_i q +  \left[1+2\Phi\right]a^2\nabla^2 \chi  - a^2 h^{ij}\partial_i\partial_j \chi+\left(
    -\delta^{\mu 0} q+ a^2\delta^{\mu i}\partial_{i}\chi
    \right)\partial_\mu\left(\Psi-3\Phi+\frac{1}{2}h\right),
\end{align}
\end{linenomath}
where $\nabla^2=\delta^{ij}\partial_i \partial_j$. We can find the equations of motion when solving for $q'$ and including the source on the right hand side
\begin{linenomath}
\begin{align} \label{eq:eom_masterequation}
    q' &= \left[1+2\Psi+2\Phi\right]a^2\nabla\chi-\left[1+2\Psi\right]a^4
    \partial_\phi V_{\text{eff}}/ 
    -2B^i\partial_i q
    -a^2h^{ij}\partial_i\partial_j\chi
    +\left(
    -\delta^{\mu 0} q+ a^2\delta^{\mu i}\partial_{i}\chi
    \right)\partial_\mu\left(\Psi-3\Phi\red{+\frac{1}{2}h}\right),\\
    \chi' &= q/a^2,\\
    \chi &\equiv \phi/v.
\end{align}
\end{linenomath}
We add these correction terms in the source of equation \eqref{eq:ELeq1} so that the acceleration $\dot q$ at first order becomes a function of the velocity $q$ as well. We also note that we have marked the final term red due to the trace of the tensor perturbation being completely negligible (also its derivatives), and comparable to machine precision, so that it will be neglected within our implementation. In fact it can be understood from the gauge conditions of the Poisson gauge \eqref{eq:gaugeconditions}. We have outputted its maximal value as a consistency check in some fiducial runs.

\subsubsection{Leapfrog}
The leapfrog solver is a second order relatively cheap symplectic solver that is energy conserving and allows one to stably evolve oscillatory solutions \citep{hairer_lubich_wanner_2003}.
For solving the system of equations \eqref{eq:ELeq2wdef} and \eqref{eq:ELeq1} we implement the leapfrog solver in a fixed timestep loop of finer timesteps that is run in within a more coarse adaptive timestep of the gevolution loop that solves for the rest of the fields. The number of finer timesteps for the asymmetron relative to the gevolution fields is a parameter set by the user. Within the asymmetron evolution loop that iterates over the lattice points, the pseudo code goes as follows
\begin{algorithmic}
\STATE $a_{n+1/2} = a_n\left(1+\mathcal{H}_n\,\diff\tau\,/\, 2\right)$
\STATE $\chi_{n+1}\,\,\,= \chi_n + q_{n+1/2}\,\diff\tau/a_{n+1/2}^2$
\STATE $a_{n+1} = a_{n+1/2}\left(1+\mathcal{H}_{n+1/2}\, \,\diff\tau\,/\, 2\right)$
\STATE $q_{n+1+1/2} \,\,\,= q_{n+1/2} + q'_{\,n+1}\, \diff \tau\, $
\end{algorithmic}
where the index $n$ refers to the step number. As a consistency check we have also implemented the option to solve the field using the first order Euler-Cromer integration, which is also symplectic. This goes as
\begin{algorithmic}
\STATE $q_{n+1} = q_n + q'_{\,n}\,\diff\tau$
\STATE $a_{n+1} = a_n\left(1+\mathcal{H}_n\,\diff\tau\right)$
\STATE $\chi_{n+1} = \chi_n + q_{n+1}\,\diff \tau\,/\, a^2_{n+1}$
\end{algorithmic}
The choice of the timestep is discussed more in appendix \ref{A:analytic_comparison}, but in general it should be made by the user by starting with a slightly large timestep, but on the physical timescale of the field, and then decreasing it until convergence.
\subsubsection{Gauss-Seidel relaxation}\label{SSS:GaussSeidel}
For better comparison with previous symmetron implementations in the literature, such as ISIS \citep{llinares_isis_2014}, we implement a constraint solver for the non-dynamic field. Similarly to them, we implement a Gauss-Seidel relaxation scheme for solving the constraint equation we get from employing the quasi-static approximation, seen in equation \eqref{eq:ELeq1} by putting $q'=0$. This looks similar to Newton's method and can be stated 
\begin{linenomath}
\begin{align}\label{eq:Newtonsmethod}
    \chi^{(\rm{i,j,k})}_{n+1} = \chi^{(\rm{i,j,k})}_n - \frac{f(\chi_n)}{\partial_{\chi_n^{(\rm{i,j,k})}}f(\chi_n)},
\end{align}
\end{linenomath}
where i,j,k signifies the cell coordinates, n subscript is the time step and $f$ is the function that we want to find the zeros of, which is equation \eqref{eq:ELeq1} with $q'=0$. This function contains the Laplacian, which also depends on the neighbour points on the lattice, but only the lattice point in question is being differentiated with respect to in the denominator of \eqref{eq:Newtonsmethod}. We then have
\begin{linenomath}
\begin{align}
    \chi_{i+1} =  \chi_i - \frac{f(\chi_i)}{-6/\diff x^2-V_{\phi\phi}(\phi_i)},
\end{align}
\end{linenomath}
which we iterate until the difference $|\chi_{i+1}-\chi_i|$ is smaller than some user defined tolerance. For $z\ge z_*$ we choose a small non-zero value for the starting guess $\chi_{\rm{start}}=10^{-4}$, since the field is expected to be confined to the origin by the stable unbroken vacuum potential;
For $z<z_*$ the starting guess is the positive expectation value of equation \eqref{eq:avev}.

\subsection{Stress-energy tensor}\label{SS:SEtensor}
We need to find the stress-energy tensor for sourcing the Einstein field equations for the metric degrees of freedom and for evaluating the coupling to matter in the asymmetron equations of motion. We also need it for the background in the case where we are averaging the energy over the box. The SE tensor is required in the Einstein frame for all of these purposes, but as seen in figure \ref{fig:FlowChart}, we evaluate it using both Einstein and Jordan frames; This is because we want to exploit the fact that the Jordan frame SE tensor has no couplings to the asymmetron field, and therefore is the familiar standard model SE tensor. The code thus first evaluates the SE tensor in the Jordan frame and then transforms it to the Einstein frame. Since the SE tensor is found by constructing a density field out of the particles in the simulation, it means that they will have to be evolved in the Jordan frame, which we do by evolving their geodesics with Jordan frame metric potentials, and is explained more closely in section \ref{SS:fifthforce}.
One can find the stress-energy tensor corresponding to the matter Lagrangian in general relativity as 
\begin{linenomath}
\begin{align} \label{eq:SEdef}
    T_{\mu\nu} = - \frac{2}{\sqrt{-g}}\frac{\delta \mathcal{L}_g}{\delta g^{\mu\nu}},
\end{align}\end{linenomath}
where $\mathcal{L}_g\equiv \sqrt{-g}\mathcal{L}$ and $g$ is the determinant of the metric. Applying this equation in the Jordan frame and then transforming it to the Einstein frame, we find
\begin{linenomath}
\begin{align}\label{eq:SEmTransform}
     T_{\,\,\nu}^{\mu\,(m)} = A^{4} \tilde T^{\mu\,(m)}_{\,\,\nu}.
\end{align}\end{linenomath}
It is worth mentioning that in the Einstein frame, the right hand side $tt$-component $\tilde  T^{t\,(m)}_{\,\,t}$ does not scale $\propto a^{-3}$ as is the case in $\Lambda$CDM. This scaling comes from the continuity equation $\tilde \nabla_\mu \tilde T^{\mu\,(m)}_{\,\,\nu}=0$, which does not hold for the Einstein frame $\nabla$. However, after transforming the covariant derivative and stress energy tensor, we get the relation (appendix D of \cite{wald_general_1984})
\begin{linenomath}
\begin{align}\label{eq:current}
    \nabla_\mu T^{\mu\, ( m)}_{\,\,\nu} =  \frac{\partial_\nu A}{A}  T^{(m)},
\end{align}\end{linenomath}
which shows the rate of energy and momentum exchange between the two sectors. We will consider this more closely in subsection \ref{A:energytransfercheck}. We note for now that a quantity which is conserved on the background level with respect to the Einstein frame covariant derivative is
\begin{linenomath}
\begin{align}\label{eq:difficultexplanation}
    \left\langle\nabla_\mu \left(A^3\tilde T^{\mu\,(m)}_{\,\,{t}}\right)\right\rangle = \left\langle\tilde T^{(m)} A^2 \partial_{t} A -\tilde T^{\mu\,(m)}_{\,\,{t}} A^2\partial_\mu A\right\rangle
    = -\left\langle A^2\tilde m \tilde v^i\partial_i A \right\rangle \simeq 0,
\end{align}\end{linenomath}
where we have skipped some steps involving the transformation of the Christoffel symbols $\Gamma ^{\alpha}_{\beta\gamma}$. Expression \eqref{eq:difficultexplanation} is true when we consider pressureless matter, which is the case for cold dark matter. $\tilde m$ and $\tilde v^i$ are respectively the constant mass and the velocity of the particle in the Jordan frame. The average is over the simulation volume, and should be approximately zero because of both isotropy and that $v\ll 1$. When we are referring to the background Jordan frame energy density in the Einstein frame as something that scales $\hat{\rho}_m\propto a^{-3}$, it is therefore really the quantity $\hat{\rho}_m = \left\langle\tilde \rho_m A^3\right\rangle$ we are referring to. This is a caveat to keep in mind, but it only matters for us when we are doing the background analysis in equation \eqref{eq:backgroundsymcontribution}.

We will add the expression for the Einstein frame matter SE tensor to the SE tensor of the asymmetron to get the total SE tensor
\begin{linenomath}
\begin{align}
    T^{\mu}_{\,\,\nu} = T^{\mu\,(\phi)}_{\,\,\nu}  +T^{\mu\,(m)}_{\,\,\nu},
\end{align}\end{linenomath}
that is conserved in the Einstein frame $ \nabla_\mu T^{\mu\nu} = 0$. Applying equation \eqref{eq:SEdef} to the free asymmetron field, we find its SE tensor
\begin{linenomath}
\begin{align} \label{Tmunu_def}
    T^{\mu\,(\phi)}_{\,\,\nu}= \partial^\mu\phi\partial_\nu\phi + \delta^\mu_\nu\mathcal{L}.
\end{align}\end{linenomath}
After some algebra, expanding to the first order in the metric potentials, we obtain 
\begin{linenomath}
\begin{align}\label{eq:fieldenergydensity}
    \rho_{(\phi)} \equiv -T^t_{\,\,t} &= \frac{1}{2}\dot \phi^2\left(1-2\Psi\right) + V + \frac{\partial_i\phi\partial_j\phi}{2a^2}\left( \delta^{ij}\left[1+2\Phi\right]-h_{ij}\right),\\
    \mathcal{P}^{(\phi)}_i \equiv T^t_{\,\,i} &= -\frac{\partial_i\phi}{a}\left(\dot\phi\left[1-2\Psi\right] + \frac{B^j}{a}\partial_j\phi\right),\\
    S^{i\,(\phi)}_{\,\,j}/2 \equiv T^i_j &= \partial_j\phi \left( - \dot\phi B^i + \partial_k\phi\left[ \delta^{ki}\left\{1+2\Phi\right\}-h^{ki}\right]\right) + \delta^i_{j}\mathcal{L},\\
    \mathcal{L} &= - \frac{1}{2}\left(
    -\left[1-2\Psi\right] \dot\phi^2
    - 2\dot\phi B^i\partial_i \phi 
    + \partial_i\phi\partial_j\phi\left\{\delta^{ij}\left[1+2\Phi\right] - h^{ij}
    \right\}
    \right) - V,
\end{align}\end{linenomath}
which we can implement in our code and compute at every time step. We have here connected the stress energy tensor to the fluid variables energy density $\rho$, momentum density $\mathcal{P}^i$ and stress tensor $S^i_{\,j}$. Furthermore, we have the isotropic pressure $p = S^i_{\,\,i}/3$.
Looking at equation \eqref{eq:SEmTransform} we see that in addition to the uncoupled Einstein frame matter stress-energy tensor, there is an interaction term when expanding in $\Delta A$
\begin{linenomath}
\begin{align}\label{eq:interactionenergy}
    T_{\,\,\nu}^{\mu\,(\phi m)} = 2\left(\frac{\phi}{M}\right)^2 \tilde T_{\,\,\nu}^{\mu\,(m)} ,
\end{align}\end{linenomath}
so we can keep track of the interaction energy separately as
\begin{linenomath}
\begin{align}
    \rho_{\phi m} = 2 \left(\frac{\phi}{M}\right)^2\tilde \rho_m.
\end{align}\end{linenomath}
We will find it instructive to consider the equation of state parameter of the field in which we do not include the interaction
\begin{linenomath}
\begin{align}\label{eq:equationofstateequation}
    \omega = \frac{p}{\rho} = -\frac{1}{3}\frac{T^i_i}{T^t_t},
\end{align}\end{linenomath}
that will be $-1$ for the background when the field is non-dynamic.
We can see this by writing out the equation above at background level, setting spatial gradients and metric perturbations to zero
\begin{linenomath}
\begin{align}\label{eq:eosIsMinus1}
    \bar\omega_{\phi} = \frac{\frac{1}{2}\dot\phi^2 - V}{\frac{1}{2}\dot \phi^2 + V},
\end{align}\end{linenomath}
which tends towards $-1$ as the kinetic energy becomes negligible in comparison to the potential (i.e. when the field is quasi-static).
If we assume quasi-static evolution, so that time-derivatives disappear, then for the background energy density, one finds
\begin{linenomath}
\begin{align}\label{eq:backgroundsymcontribution}
    \bar\rho_\phi + \bar\rho_{\phi m} = V \left(\bar\phi\right) + \frac{1}{2}\left( \frac{\bar\phi}{M}\right)^2  \Omega_{m,0}\,\rho_{c,0}/a^3,
\end{align}\end{linenomath}
while the background field's expectation value, $\bar\phi$, is found using \eqref{eq:potentialZeros}. The factor $1/2$ instead of $2$ as in equation \eqref{eq:interactionenergy} follows from the argument we made above, in equation \eqref{eq:difficultexplanation}. We insert this into the background Friedmann equation in equation \eqref{eq:symbackground} for comparison with the box-averaged energy densities. We only apply this analytic approach to the symmetron case, as it is not clear how to weight the two different minima post symmetry breaking for the asymmetron, although a good guess assuming unstable domain walls can be to weigh the deepest minimum 100 $\%$.

\subsection{Fifth force}\label{SS:fifthforce}
Following the procedure performed in \cite{adamek_gevolution_2016}, we find the geodesic equations for the matter particles from the particle Lagrangian
\begin{linenomath}
\begin{align}
    L_p = -m\int \diff \tau \sqrt{g_{\mu\nu}\frac{\diff x^\mu}{\diff \tau} \frac{\diff x^\nu}{\diff \tau}  },
\end{align}\end{linenomath}
where  $x^\mu$ denotes the coordinate of the particle, and $\tau=\int \diff t/a$ is the conformal time that we use in the simulation. A way to include the fifth force in the Einstein frame is to realise that the geodesic equation gives the complete description for the movement of the particles in the Jordan frame, and that we can transform quantities from there to the Einstein frame by the conformal transformation \eqref{eq:conformaltransform}. Since the results of \cite{adamek_gevolution_2016} are valid when expressed in the Jordan frame metric potentials, we express those equations with the Einstein frame potentials transformed to the Jordan frame using equations \eqref{eq:metrictransform1} to \eqref{eq:metrictransform-1}.
In gevolution the metric is linearised in a smallness parameter $\epsilon$, equation \eqref{eq:epsilon}, which gives us the constraint \eqref{eq:symmetronGeodesicConstraint} for $\Delta A$. If this constraint is satisfied, the higher order terms can be dropped and we find the coordinate three-velocity with respect to the conformal time of the particle using the Einstein frame potentials\footnote{Note that we write $v_i\equiv v^i\delta_{ij}$ since we do not use it with index downstairs. Indices are treated the same way for the perturbations.} 
\begin{linenomath}
\begin{align}\label{eq:vofq}
    v_i = \frac{q_i}{\sqrt{q^2+m^2 a^2}}\left\{
    1+ \Psi + \frac{1}{2}\left(\frac{\phi}{M}\right)^2
    + \left( 2 - \frac{q^2}{q^2+m^2a^2}    \right) \left[\Phi-\frac{1}{2}\left(\frac{\phi}{M}\right)^2\right] \red{+ \frac{1}{2}\frac{h_{jk}q^jq^k}{\sqrt{q^2+m^2 a^2}}}
    \right\} +
    B_i \red{-  \frac{h_{ij}q^j}{\sqrt{q^2+m^2 a^2}}},
\end{align}\end{linenomath}
and the conformal acceleration of the canonical momentum
\begin{linenomath}
\begin{align} \label{eq:conjacc}
    \frac{\diff q_i}{\diff \tau}
    = - \sqrt{q^2+m^2 a^2}\left\{ 
    \Psi_{,i} + \frac{\phi}{M^2}\partial_i\phi + \frac{q^2}{q^2+m^2a^2}\left(\Phi_{,i}-\frac{\phi}{M^2}\partial_i\phi\right)
    + \frac{q^j B_{j,i}}{\sqrt{q^2+m^2a^2}}
    \red{-\frac{1}{2}\frac{q^j q^k h_{jk,i}}{q^2+m^2a^2}}
    \right\}.
\end{align}\end{linenomath}
We see that when in the Newtonian limit $v \sim q/m\ll 1$, $q_i = ma v_i$, and neglecting the Hubble factor, we find $\diff q_i/\diff\tau = m v'_i = -ma (\partial_i\Psi+\frac{\phi}{M^2}\partial_i\phi)$, which is what we expected in equation \eqref{eq:fifthforce}. For our non-relativistic implementation, we therefore use
\begin{linenomath}
\begin{align}
    \frac{\diff v_i}{\diff t} =  -\partial_i\Psi - \frac{\phi}{M^2}\partial_i\phi.
\end{align}\end{linenomath}
Note that the last terms of equations \eqref{eq:vofq} and \eqref{eq:conjacc} are red. This is to indicate that they are not included in the gevolution implementation of \cite{adamek_gevolution_2016} due to the claim that the effect of scattering of particles from the $h_{ij}$ field is negligible in the standard model of cosmology. We will therefore neglect these terms ourselves here for now, because we do not have any reason to expect this to change in the asymmetron model.

We implement the new terms in equations \eqref{eq:vofq} and \eqref{eq:conjacc} straightforwardly among the rest that are already implemented in gevolution for the movement of baryons, cold dark matter and relativistic particles. First, the conjugate canonical momentum is evolved using equation \eqref{eq:conjacc}, and then the particles' positions are updated using the velocity found with equation \eqref{eq:vofq}.The stress energy tensor is then constructed using gevolution's routine that includes geometric corrections, explained in \cite{adamek_gevolution_2016}, dependent on the Jordan frame metric potentials. The SE tensor is afterwards multiplied by $A^4$ to get the Einstein frame SE tensor. At last, the Einstein frame SE tensor is used for the Einstein field equations to solve for the Einstein frame metric potentials, see figure \ref{fig:FlowChart}.

\section{Results and discussions} \label{S:results}

In this section we will first review some of the tests that were made in order to validate the code, and how the different implementation options compare, and then move on to discussing some of the interesting features of the model as seen through the simulation output. In all of the simulations, for consistency, we have fixed the cosmology $(h,\Omega_m,\Omega_b,\Omega_\Lambda) = (0.719,0.279,0.0437,0.72)$, where $h$ is the Hubble parameter $H_0/100$ and $\Omega_m,\Omega_b,\Omega_\Lambda$ are the total matter, baryonic and vacuum cosmological energy density parameters respectively. The choice of cosmology is made for better comparison with \cite{llinares_isis_2014}\footnote{We thank the authors of \citep{llinares_isis_2014} for clearing up a mismatch with our power spectrum and theirs, owing to a small error in their paper. Our figure agrees with the power spectrum that is presented in the earlier ISIS paper \cite{davis_structure_2012}.}\footnote{Note that our parameters are slightly off that of \cite{llinares_isis_2014} since we had to specify the baryonic fraction for the CLASS run and therefore chose something closer to Planck cosmology for the matter density parameter $\Omega_m/h^2$.}, and otherwise the parameters are Planck cosmology \citep{planck_planck_2020}. 

\subsection{Code validation} \label{SS:codevalidation}
\begin{figure}
    \centering
    \includegraphics[width=0.8\linewidth]{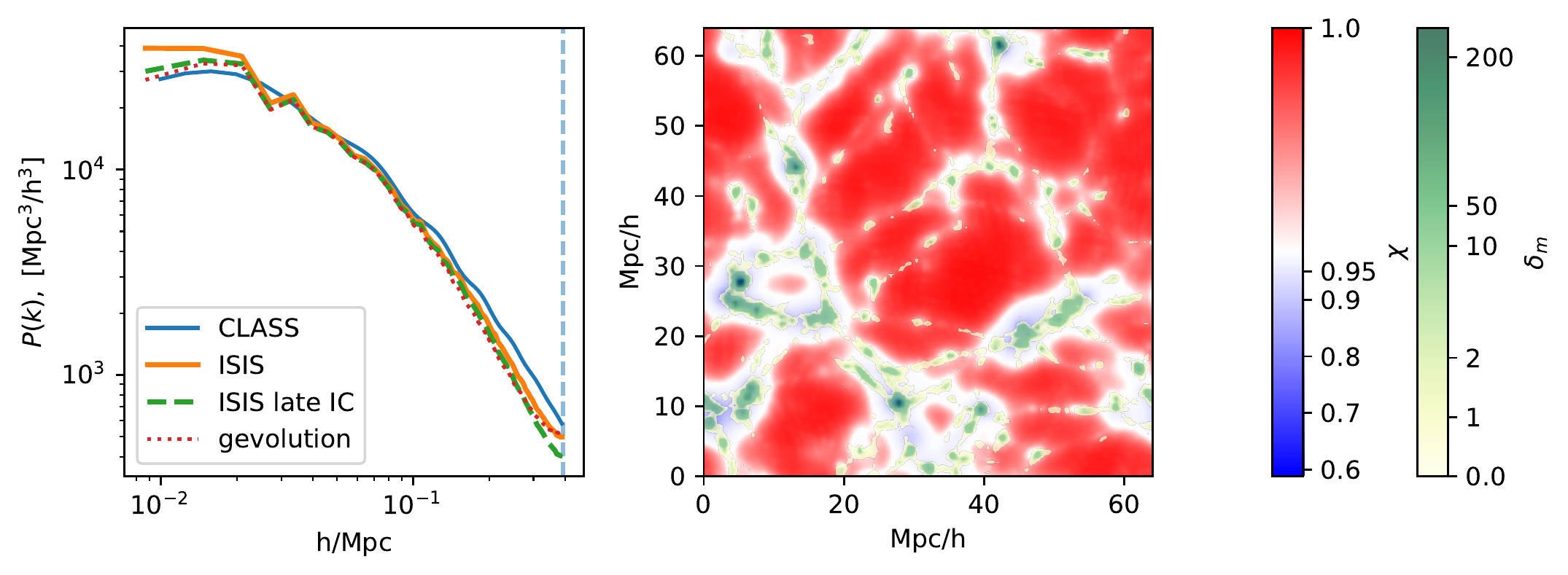}
    \caption{Left: The $\Lambda$CDM matter power spectra compared between gevolution, CLASS and ISIS at redshift $z=0$, when run with the same initial conditions. The spectrum labelled `ISIS late IC' is initialised at redshift $z=30$ instead of $z=100$. Right: Slice of the simulation volume, displaying the matter overdensity field (only $\delta_m>0$) overlayed the scalar field background for the asevolution quasi-static approximation run, showing a large degree of correlation with the cosmic structures.}
    \label{fig:sanitycheck_isis}
\end{figure}
We compare the results from asevolution with the ISIS code.
ISIS \citep{llinares_isis_2014} is a modification of the $N$-body solver RAMSES \citep{teyssier_cosmological_2002} that solves for several different modified gravity theories, symmetron, $f(R)$ and DGP, assuming the quasi-static approximation. We compare our results to similar runs performed with the ISIS code as a benchmark test. To that end, we have implemented a similar constraint solver for the scalar field, making use of the quasi-static approximation, so that we can untangle which differences are due to the different implementations of gevolution and RAMSES, and which are due to e.g. the dynamics we are now solving for. For the same reason, we have chosen to turn off the effect of the symmetron on the background and all relativistic effects on the symmetron, as we are comparing with the ISIS that has neither of these.
We use the same initial conditions for both codes which are generated by CLASS transfer functions at redshift $z=100$. The symmetron parameters used in the comparison are $(\xi_*,a_*,\beta_*) = (3.3\cdot 10^{-4},0.33,1)$, since they are also used in \cite{llinares_isis_2014}. \oen{The choice of parameters can be motivated from the following: we choose $a_*\sim 1$ from the fact that we want a late-time symmetry breaking that evades early universe constraints and allows for large symmetron domains according to \cite{llinares_domain_2014}; we set $\beta_*=1$ since the most natural choice for force strength relative to gravity of another force acting on cosmological scales, is $\sim 1$; and we choose $L_C \sim 1$ Mpc/h since for this choice of parameters the Compton wavelength is according to \cite{hinterbichler_symmetron_2010} constrained to be $\lesssim 0.5$ Mpc/h, which we are exceeding slightly}. As in the ISIS paper, we are using\footnote{NB: these are units defined in subsection \ref{SS:units}: $h^{-1}$cMpc means that the distance is in comoving Mpc/h.} $(64\,\,\,h^{-1}\text{cMpc})^3$ volume boxes and $128^3$ particles and grids.

\begin{figure}
    \centering
    \includegraphics[width=0.6\linewidth]{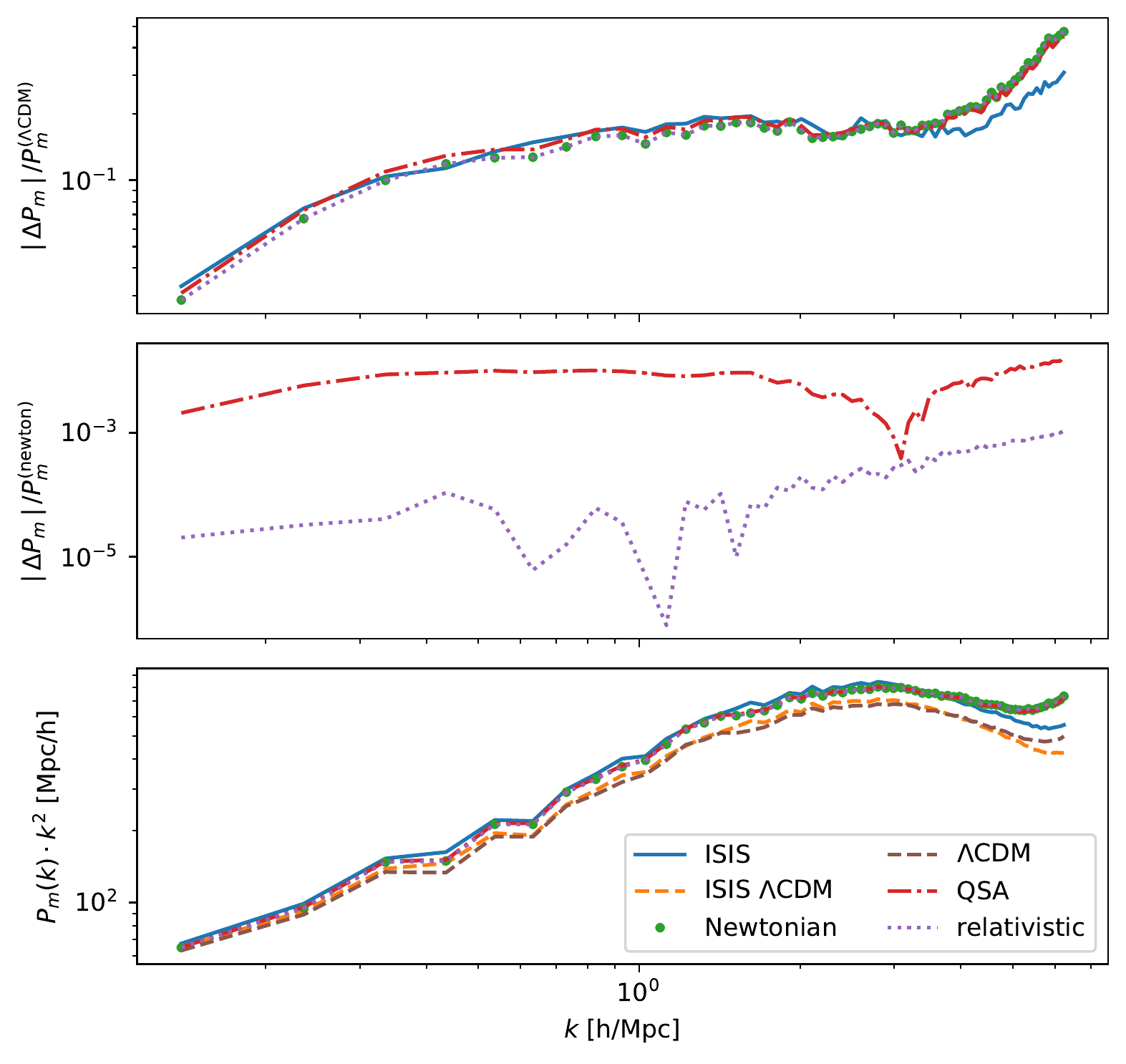}
    \caption{Legend: All simulations apart from the ones specified ISIS are made in asevolution. All simulations not specified $\Lambda$CDM are for the symmetron. Top panel: The relative difference between the symmetron and $\Lambda$CDM power spectra for the different simulation choices. The comparison is to the $\Lambda$CDM simulation of the respective code. Middle: Relative power spectra to the non-relativistic dynamic implementation. Bottom: The matter power spectra in units Mpc/h.
    The power spectrum that is presented with the same parameter choices as in \cite{llinares_isis_2014} is shown in solid blue. Symmetron parameters are here $(\xi_*,a_*,\beta_*) = (3.3\cdot 10^{-4},0.33,1)$. 
  }
    \label{fig:powerspectrum_comparison_isis}
\end{figure}

\begin{figure}
    \centering
    \includegraphics[width=0.7\linewidth]{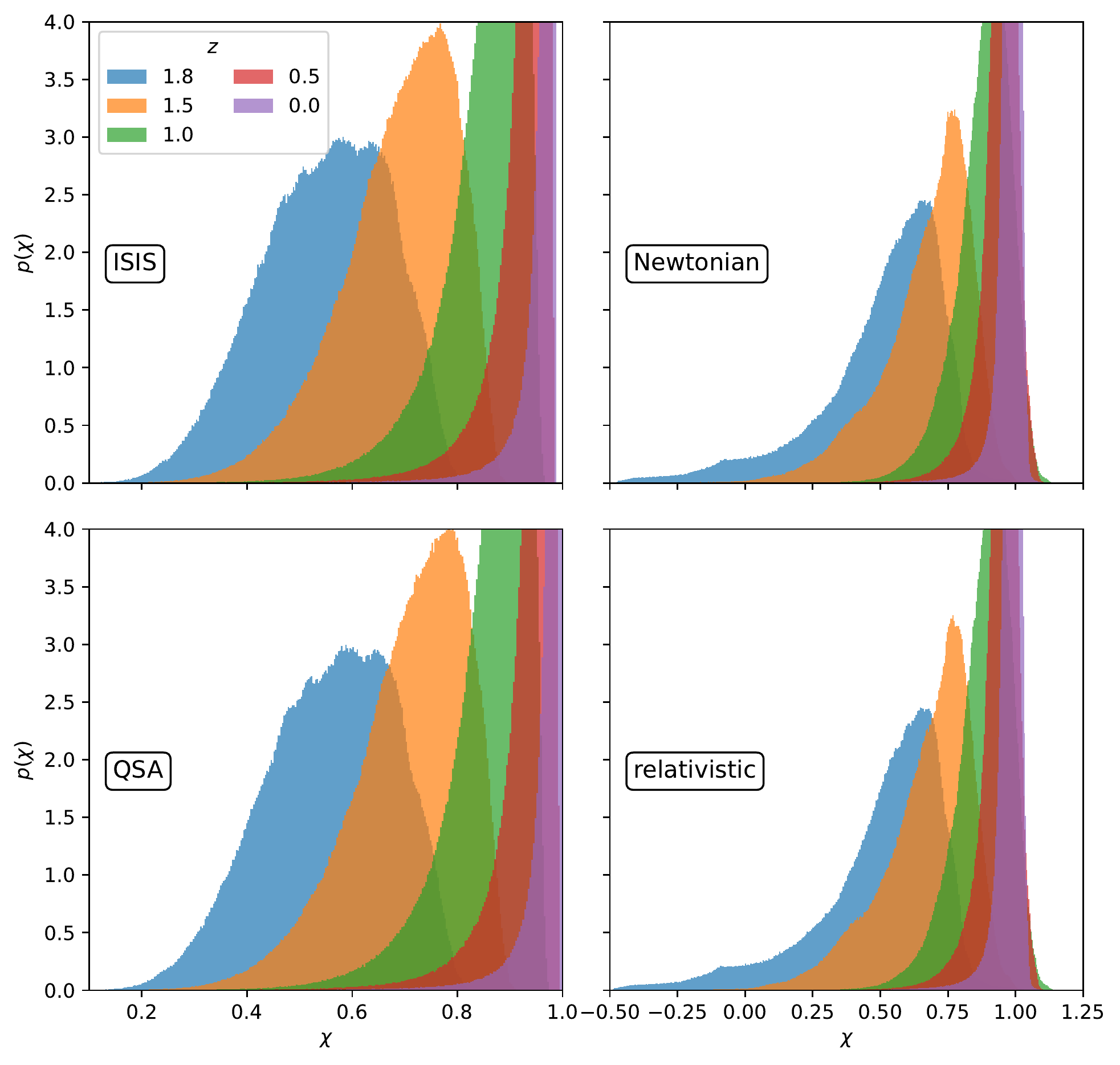}
    \caption{Corresponding probability densities, $p(\chi)$, of the dimensionless symmetron field, $\chi$, in a $(64\,\,\,h^{-1}$cMpc$)^3$ box as provided by (from top left to bottom right) (1) ISIS and then (2) asevolution with Newtonian geodesics and dynamic symmetron, (3) asevolution with Newtonian geodesics and quasi static approximation for symmetron and finally (4) asevolution with dynamic symmetron and relativistic geodesics. The redshift of the histogram is indicated in the legend.
    Symmetron parameters are here $(\xi_*,a_*,\beta_*) = (3.3\cdot 10^{-4},0.33,1)$. 
    }
    \label{fig:histograms_comparison_isis}
\end{figure}

\begin{figure}
    \centering
    \includegraphics[width=0.7\linewidth]{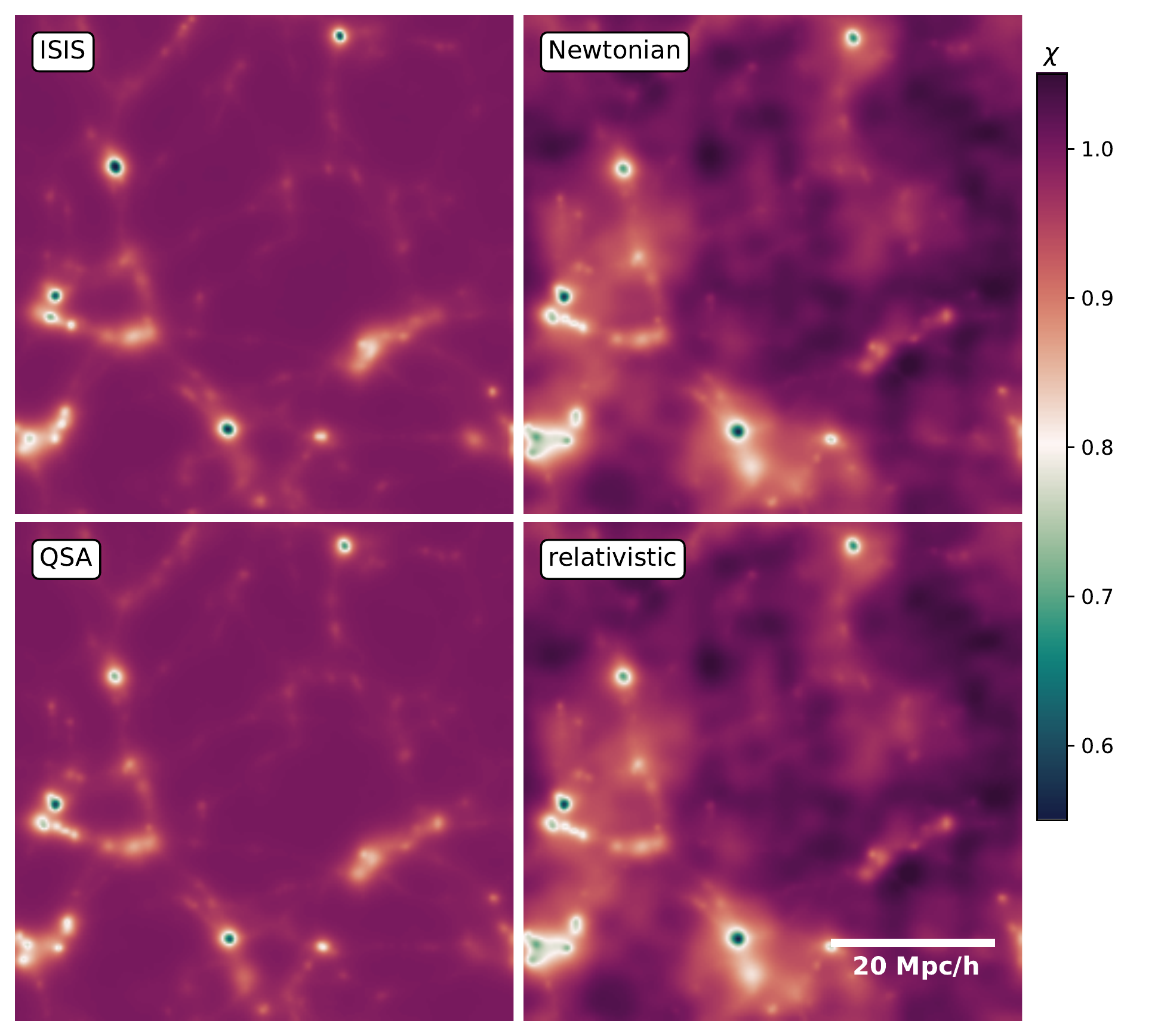}
    \caption{Corresponding slices of the dimensionless symmetron field, $\chi$, at redshift $z=0$ in a $(64\,\,\,h^{-1}\text{cMpc})^3$ box as provided by (from top left to bottom right) (1) ISIS and then (2) asevolution with dynamic symmetron and Newtonian geodesic, (3) asevolution with Newtonian geodesic and quasi static approximation and finally (4) asevolution with dynamic symmetron and relativistic geodesic. The redshift of the histogram is indicated in the legend. Symmetry breaking happens here at $z_*=2$.
    Symmetron parameters are here $(\xi_*,a_*,\beta_*) = (3.3\cdot 10^{-4},0.33,1)$.
    }
    \label{fig:slice_comparison_isis}
\end{figure}

At first we found a small mismatch between the result of RAMSES and gevolution in the enhancement of the dark matter power spectrum at large scales and $z=0$, but we see a similar disagreement in their $\Lambda$CDM power spectra in the left of figure \ref{fig:sanitycheck_isis}, which one can see we managed to improve by initialising RAMSES at a later redshift, here $z = 30$. Using this initialisation, which is of no consequence with regards to the symmetron since symmetry breaking does not take place until redshift $z_* = 2$, we find a good agreement between the two simulations, as shown in figure \ref{fig:powerspectrum_comparison_isis}. Furthermore, the histograms in figure \ref{fig:histograms_comparison_isis} and the snapshots in figure \ref{fig:slice_comparison_isis} show a near identical field configuration of the asevolution quasi-static implementation and that of ISIS. One can see a small horizontal relative displacement of the final redshift histogram between ISIS and the quasi-static solver, which we found to be due to ISIS outputting less exactly at the requested redshift than asevolution, missing here by a difference $\Delta z\sim 0.001$. \oen{The disagreement with CLASS on small- and large scales we presume to be owing to the coarse simulation, the non-linear corrections on small scales and cosmic variance on large scales}. Finally, as a sanity check, we see a perfect correlation between the dark matter structure and the screening fraction of the symmetron field in the right side of figure \ref{fig:sanitycheck_isis}, where we have overlayed the overdensity field and the symmetron field snapshots. This agrees with what we expect from our discussion of the screening mechanism in section \ref{S:theory}. We notice that the field is not completely screened, but instead tends towards $\chi=0.5$ in high density environments where we expect it to be screened; This is owing to the low resolution of the simulation, and we see a complete screening $\chi=0$ for improved resolution simulations. \oen{The fact the field only can be seen to take the positive minimum in the right side of figure \ref{fig:sanitycheck_isis} can be understood from the fact that the initial guess of the Gauss-Seidel relaxation, see section \ref{SSS:GaussSeidel}, is chosen only to be the positive minimum of equation \eqref{eq:expectation_sym}; When using the quasi-static approximation, we are loosing out on the phenomenology of domain walls that we don't put in by hand ourselves.}

In figure \ref{fig:powerspectrum_comparison_isis}, we have considered different simulation cases for the symmetron field: 1) QSA: using the quasi static approximation for the symmetron field in asevolution, 2) $\Lambda$CDM: $\Lambda$CDM case in asevolution, 3) Newtonian: considering dynamic symmetron field in asevolution using the Newtonian geodesic equation and the non-relativistic field equations \eqref{eq:ELeq1}, 4) relativistic: considering dynamic symmetron field in asevolution with relativistic geodesic equations and field equations \eqref{eq:eom_masterequation}. We see a $10\%$ difference on the power spectrum with respect to $\Lambda$CDM at as large scales as $k\sim 0.03$ h/Mpc. We find a small and negligible effect on the enhancement of the dark matter power spectrum of including dynamics or relativistic degrees of freedom for our choice of parameters, as seen in figure \ref{fig:powerspectrum_comparison_isis}.  However, although small, we do find a visible impact on the configuration of the field from including its dynamics. It is visible in figure \ref{fig:slice_comparison_isis} as a `smokey' contour about the configuration indicated by the quasi-static solution. It is also visible in the right-side histograms of values of the field across the lattice of the box in figure \ref{fig:histograms_comparison_isis} where we see a more slow evolution of the field away from the origin to its minimum, and additionally the acquisition of negative field values as the field initially falls into either of the two minima in different locations. Again, the relativistic effects seem to be negligible. In the middle panel of figure \ref{fig:powerspectrum_comparison_isis} we see that the relativistic effects on the matter power spectrum are smaller than the one of including dynamics by a relative difference of 1--2 orders of magnitude. We do however expect relativistic effects to be more important on large scales. Since this simulation volume is rather small, the small observed relativistic effects here are not surprising.

\subsection{Background}\label{SS:backgroundresults}
\begin{figure}
    \centering
    \includegraphics[width=0.6\linewidth]{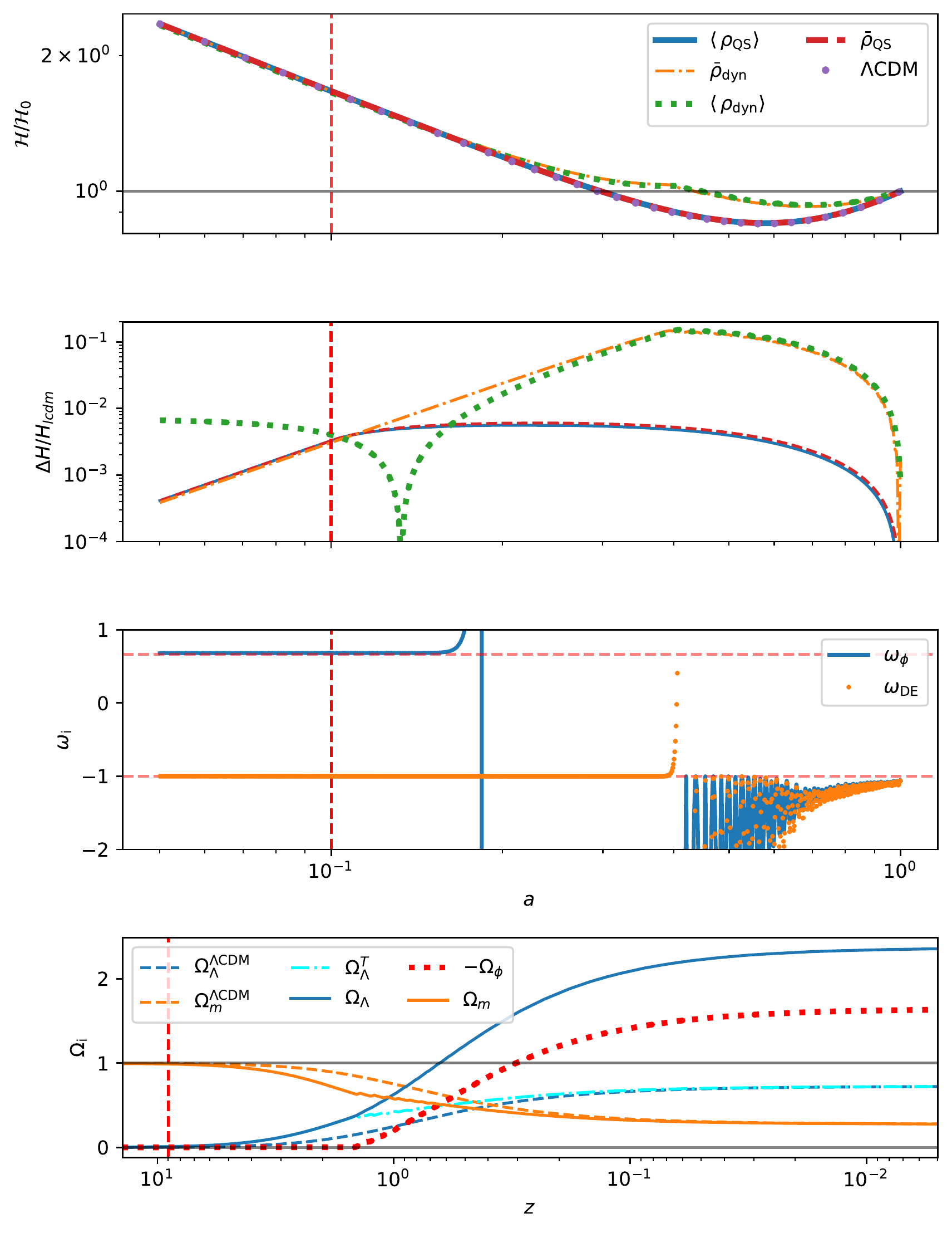}
    \caption{
    Top: Evolution of the conformal Hubble factor with respect to the scale factor. In the legend, $\rho$ indicates which energy density was used in the Friedmann equation. $\langle \rho\rangle$ means that the energy was found exactly and then averaged over the box, whereas by $\bar \rho$ it is meant that the analytic background has been used. `dyn' in subscript means that the field is  dynamic, and by `QS' that it is quasi-static. Second figure: The same background quantities are shown as relative differences with respect to the $\Lambda$CDM.
    Third plot: Box-averaged equation of state parameters for the asymmetron field, $\omega_{\phi}$, and the total dark energy, $\omega_{\rm{DE}}$, field as a function of scale factor. Horizontal dashed line indicates $\omega=-1,2/3$ respectively. Bottom row: Cosmological energy density parameters for the different sectors as a function of redshift $z$. $\Omega_\Lambda^T$ is the total dark energy, including the asymmetron. Dashed lines indicate the $\Lambda$CDM values. $\Omega_{\phi}A$ is the asymmetron part. The vertical red dashed line indicates the symmetry breaking in every subfigure.}
    \label{fig:background_backreaction}
\end{figure}
\begin{figure}
    \centering
    \includegraphics[width=0.6\linewidth]{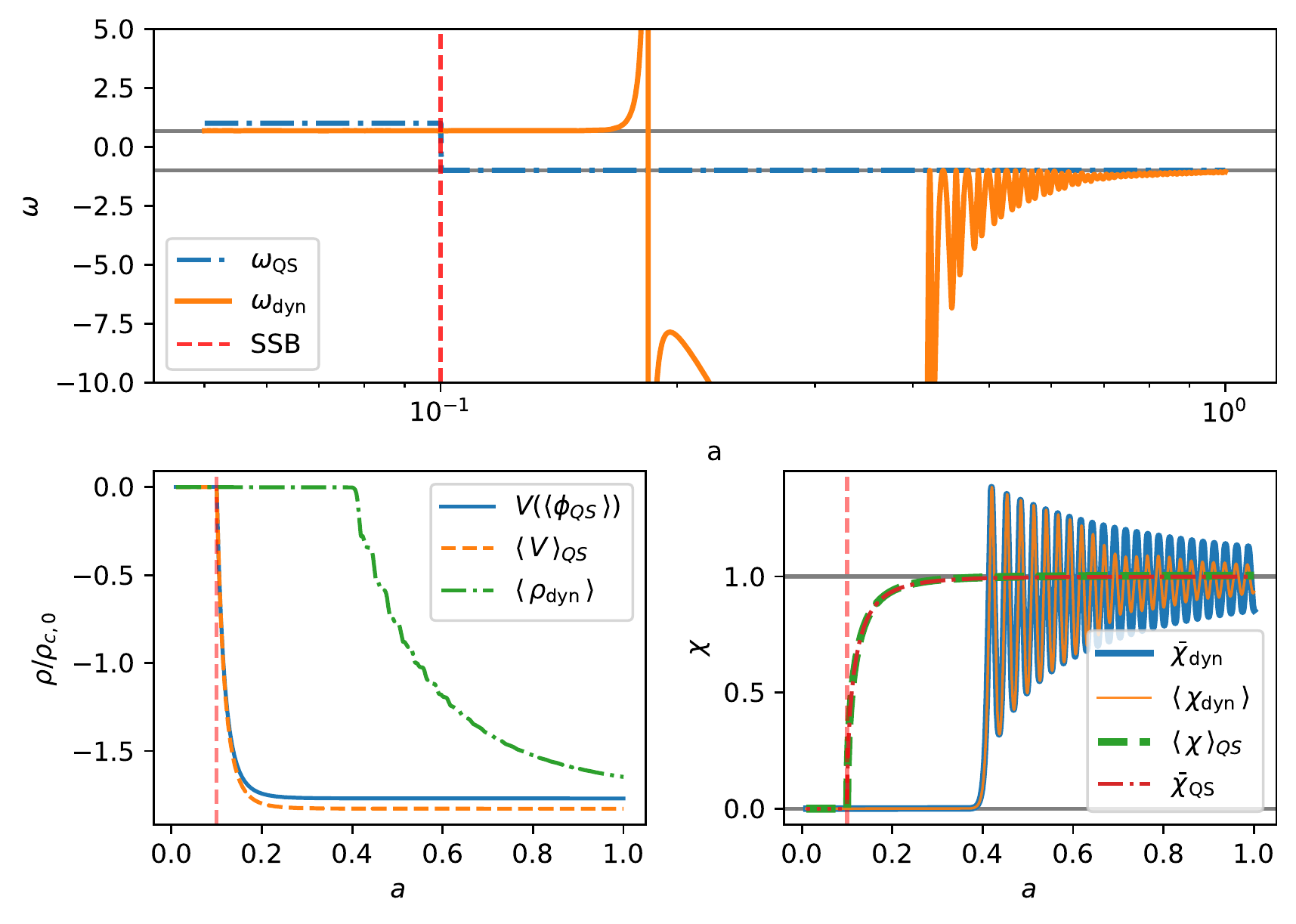}
    \caption{
    Top: Box-averaged equation of state parameters for the quasi-static and dynamic runs, as a function of the scale factor. Horizontal lines indicate $\omega=2/3,-1$ respectively. $\omega_{QS}=1$ before SSB.
    The symmetron parameters here are $(\xi_*,a_*,\beta_*)=(0.004,0.1,1)$ and the scale factor at symmetry breaking is indicated with the vertical line. Bottom left: Energy densities normalised to present time critical density, $\rho/\rho_{c,0}$, as a function of the scale factor. We plot the potential as a function of the averaged quasi-static scalar field $V(\langle\phi_{\rm{QS}}\rangle)$, the averaged potential across the box $\langle V \rangle_{\rm{QS}}$ for the quasi-static field, and the average energy density of the dynamic field, $\langle\rho_{\rm{dyn}}\rangle$. Bottom right: The expected background values, $\bar \chi_{\rm{QS}},\bar \chi_{\rm{dyn}}$ as a function of scale factor, and overlayed the box-averaged scalar field value across the box, for the quasi-static, $\langle \chi\rangle_{\rm{QS}}$, and dynamic, $\langle \chi_{\rm{dyn}}\rangle$, scalar fields.}
    \label{fig:symmetron_background}
\end{figure}
\noindent
Considering the evolution of the background, we compare five different implementations in figure \ref{fig:background_backreaction}. Apart from the $\Lambda$CDM background, where we used equation \eqref{eq:friedmann} and \eqref{eq:symbackground} with $\bar\phi=0$, we considered: 1) using the analytic solution of the background coming from solving the Friedmann equation \eqref{eq:friedmann} with $\bar\rho_{\rm{QS}}$ from equation \eqref{eq:symbackground} expressed with the quasi-static field $\bar \phi_{\rm{QS}}$ defined in equation \eqref{eq:expectation_sym}; 2) using $\bar\rho_{\rm{dyn}}$ from equation \eqref{eq:symbackground} without fixing the field value to its vacuum expectation but instead considering the dynamic $\bar\phi_{\rm{dyn}}$, found from evolving equation \eqref{eq:tmp1} at background level; or using the energy density of the field $\phi(x)${, equation \eqref{eq:fieldenergydensity},} averaged over the box for either 3) the case of the quasi-static assumption, $\langle\rho_{\rm{QS}}\rangle$, where $\dot\phi\sim 0$, or 4) for the dynamic case, $\langle\rho_{\rm{dyn}}\rangle$, where the field is evolved in time. The vacuum energy density is in each case chosen to satisfy the flatness condition that $\sum_i\Omega_i = 1$. For simplicity, and with our previous discussion in section \ref{SS:codevalidation} in mind, we choose to turn off all relativistic corrections to the symmetron in this analysis. We show the result of the comparison in figure \ref{fig:background_backreaction} for the case of symmetron parameters $(\xi_*,a_*,\beta_*) = (0.004,0.1,1)$, where the Compton wavelength $L_c \sim 12$ Mpc/h is chosen to be large on purpose so that the field is light enough to act dynamically on cosmological timescales. This parameter choice is not necessarily consistent with experiments since Compton wavelengths $L_C>1$ Mpc/h are constrained by local solar system experiments according to \cite{hinterbichler_symmetron_2010}. We are considering this scenario for testing the code where the dynamic behaviour of the field and its effect on the background is more prominent. This is pushing our perturbation in $\Delta A = A-1$ to its limit since now $\Delta A \sim 10^{-2}$. We have here also turned off the fifth force, since the condition in equation \eqref{eq:symmetronGeodesicConstraint} is not valid. However, this condition is only necessary for including the fifth force effects on the particles' geodesics, and should not\footnote{Except for a momentum transfer which should be a correction of order $\mathcal{O}(v^2)\sim 10^{-6}.$} affect the box-averaged energy density; $\tilde a^3\tilde \rho$ should be approximately constant on the background, see equation \eqref{eq:difficultexplanation}. For the energy density averaging simulations, we have used boxes with $(1000\,\,\,h^{-1}\text{cMpc})^3$ volume and $128^3$ number of particles and grids. There is found a perfect agreement between the symmetron analytic background of equation \eqref{eq:backgroundsymcontribution} and the one found from averaging the energy density across the box for the quasi-static approximation; They differ from the $\Lambda$CDM background at percent level. The agreement between the dynamically evolved background and the dynamic average options is also very well. The inclusion of dynamics lifts the discrepancy with the $\Lambda$CDM background towards $10\%$, and gives a slightly oscillating Hubble parameter in the late-time, as the field oscillates and settles into its potential minima. We understand this by looking at the top subfigure of figure \ref{fig:symmetron_background}, where we plot the box-averaged equation of state parameter of the field, which indicates whether or not the field is settled into a minimum, see equation \eqref{eq:eosIsMinus1}. Similarly, we see this in the bottom right subfigure of figure \ref{fig:symmetron_background}, which displays the average field-value. Around symmetry breaking as the minima at each point across the box moves from the origin, the field is staying at the origin momentarily, but eventually is accelerated to fall into a non-zero minimum and settles into it. The delay can be understood from equation \eqref{eq:tmp1} and that the Hubble diffusion term still dominates at the time of symmetry breaking for small masses $\mu$. The equation of state parameter of the quasi-static field shows a clear step-function behaviour around symmetry breaking; This would indicate that the quasi-static scalar field immediately starts acting like a cosmological constant and adds to the vacuum energy density of $\Omega_\Lambda$. The same argument applies to the analytic quasi-static background which by construction has the equation of state parameter $\omega_{\phi}=-1$. We might not expect the quasi-static energy-averaging approach to agree perfectly on the background with the analytic symmetron background since their energy densities differ in the bottom left subfigure of figure \ref{fig:symmetron_background}, which we can understand from the potential being non-linear
\begin{linenomath}
\begin{equation}\label{eq:non-linearpotentialcavaet}
    V(\langle \phi\rangle) \neq \langle V(\phi) \rangle.
\end{equation}\end{linenomath}
where by $\langle \rangle$ we mean spatial average over the lattice in the simulation box. We find good agreement with the analytic background value $\phi_b = \langle \phi \rangle$. However, the difference of $V(\langle \phi\rangle)$ and $\langle V(\phi) \rangle$ \eqref{eq:non-linearpotentialcavaet} seems to only cause a displacement of a constant term, whose effect is removed by our choosing the vacuum energy density parameter $\Omega_\Lambda$ in each case to satisfy the flatness condition. Finally, we also show in the bottom right of figure \ref{fig:symmetron_background} that the dynamic field has a much longer transition period, which would explain the mismatch in the Hubble calculations of figure \ref{fig:background_backreaction}, though it does approach the same asymptotic potential value as in the quasi-static case.

In the two bottom subfigures of figure \ref{fig:background_backreaction} some more information about the cosmological background is given. Here we demonstrate the time evolution of the different cosmological energy density parameters $\Omega_i=\rho_i/\rho_c$, comparing with the $\Lambda$CDM scenario. We see a slightly smaller fractional density for the matter energy density parameter in the symmetron scenario shortly after symmetry breaking. The total dark energy fractional energy density approaches that of the $\Lambda$CDM one in the late time. The total equation of state of the dark energy field $\omega_{\rm{DE}}$ is close to $-1$ for most of the time, apart from nearby the symmetry breaking, where it transitions first to large values, and then approaches $-1$ from below. This transition with a divergence in $\omega$ is similar to what is described in figure 3 of \cite{tsujikawa_constraints_2008}.

\subsection{Symmetry breaking}\label{SS:symmetrybreaking}

\begin{figure}
    \centering
    \includegraphics[width=\linewidth]{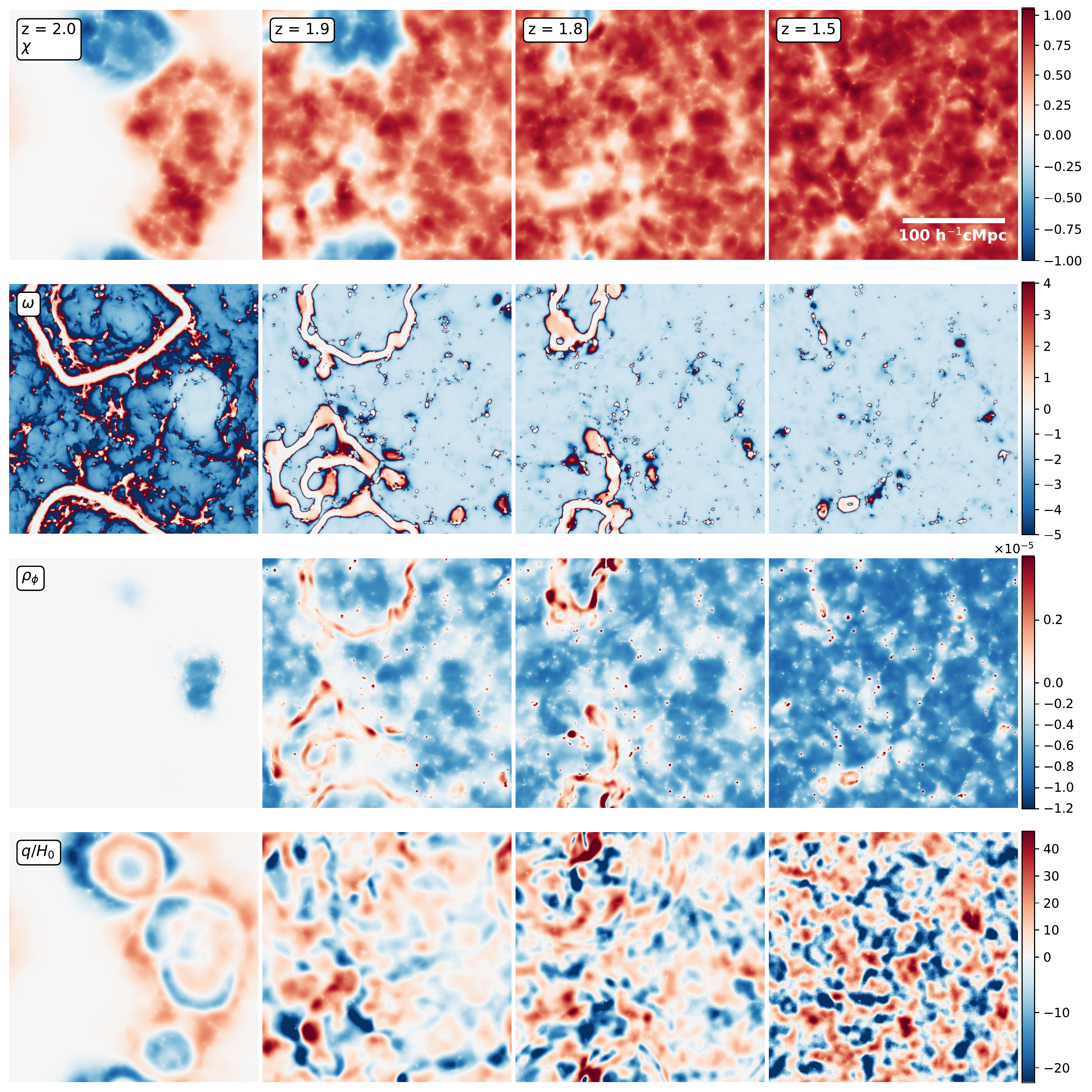}
    \caption{The process of symmetry breaking illustrated through cross-section snapshots of the $(256\,\,\,h^{-1}\text{cMpc})^3$ simulation volume, showing the normalised scalar field $\chi$, the equation of state parameter $\omega$, the free-field energy $\rho_\phi$ which is normalised to the present time critical density, and the velocity of the scalar field $q=a^2\chi'$ normalised to the present time Hubble horizon $H_0$, row-wise. Both $\omega$ and $q/H$ saturates the colour scales in isolated regions where in the most extreme cases $\omega\in(-2\cdot 10^5,6 \cdot 10^5)$ and $q/H_0\in(-70,100)$. Redshift of column is indicated in topmost row. Model shown here is $(\xi_*,a_*,\beta_*)=(3.3\cdot 10^{-4}, 0.33, 1)$.
    }
    \label{fig:SSB_cartoon}
\end{figure}

Previously, we presented the histograms in figure \ref{fig:histograms_comparison_isis} for the comparison with the ISIS code. There it can be seen that in the dynamic case, the scalar field initially falls into both minima, though eventually, by redshift $z=1.5$, it has moved entirely to the positive minimum. To study this process better, we produce a $(256\,\,\,h^{-1}\text{cMpc})^3$ volume simulation that is presented in figure \ref{fig:SSB_cartoon}. In this simulation, where the Compton wavelength also is $\sim 1 $ Mpc/h, at first, at symmetry breaking at redshift $z = 2$, there seems to be a formation of two large domains of differently signed minima bordered by a domain wall of thickness maybe $\sim 10$ times the Compton wavelength. It is interesting to note that these initial domains are far larger than the Compton wavelength of the field and should produce large-scale signatures in clustering. However, it turns out to not be stable in our simulation, and dissipate away by redshift $z = 1.5$. As stated in \cite{perivolaropoulos_gravitational_2022,derrick_comments_1964} the domain walls are generically expected to be unstable, but might be stabilised by the interaction with matter. Figure \ref{fig:SSB_cartoon} displays the slices of the simulation box for a choice of 4 different redshifts after symmetry breaking, for 4 different fields, including the equation of state parameter and the energy density of the field. The equation of state parameter is evaluated exactly at each node in the box, considering gradients and time-derivatives of the field, which allows it to be different from  $-1$, although it can be seen to take $\omega=-1$ away from the domain walls. In figure \ref{fig:background_backreaction} we see that the background equation of state parameter has the expected limits of $\omega\sim 2/3,-1$ for a free non-relativistic massive particle \citep{andersen_2012} and a total energy dominated by the potential respectively. The equation of state parameter clearly correlates spatially with the domain walls and seems to be exactly 0 for a broad band around the domains. It  then takes on large values at the border of the domain wall and the exterior/interior, where we expect large gradients of the field. Well-within the domains it approaches $-1$ as we expect. In the third row of the figure, the energy density of the field elaborates on the behaviour of the domain walls, and shows the field to have positive energy within the strip inside of the domain wall. Since the field is close to zero there, the energy contribution is mainly kinematic. The energy then becomes close to zero in two small strips that border the interior/exterior. Well within the domains the field has negative energy densities and are closer to its symmetry broken minima. The velocity of the field $q$ shows the initial fall towards the minima and outwards growth of some isolated domains. The field seems to have concentric rings of positive and negative time derivatives. In the next time steps concentric ripples are barely visible, releasing the energy of some of the domain walls. $q$ seems in general only weakly correlated with the domain walls, as we expect since the field's coherence scale should be similar to the Compton wavelength which is $1$ Mpc/h.

For the thickness, $\delta\ell$, of the domain walls we have an analytic expectation presented in \cite{llinares_domain_2014}
\begin{linenomath}
\begin{align}
    \delta\ell = \sqrt{2}L_C\left(
    1 - \frac{a_*^3}{a^3}
    \right)^{-1/2}.
\end{align}\end{linenomath}
For our parameter choice, this gives $\delta\ell(z=2.)\sim 20\,$ $h^{-1}$cMpc and for $z=1.9,1.8$ we have $\delta\ell\sim 10\,\,\,h^{-1}$cMpc. This agrees more or less with what is seen in figure \ref{fig:SSB_cartoon}.

\subsection{The asymmetron}

\begin{figure}
    \centering
    \includegraphics[width=\linewidth]{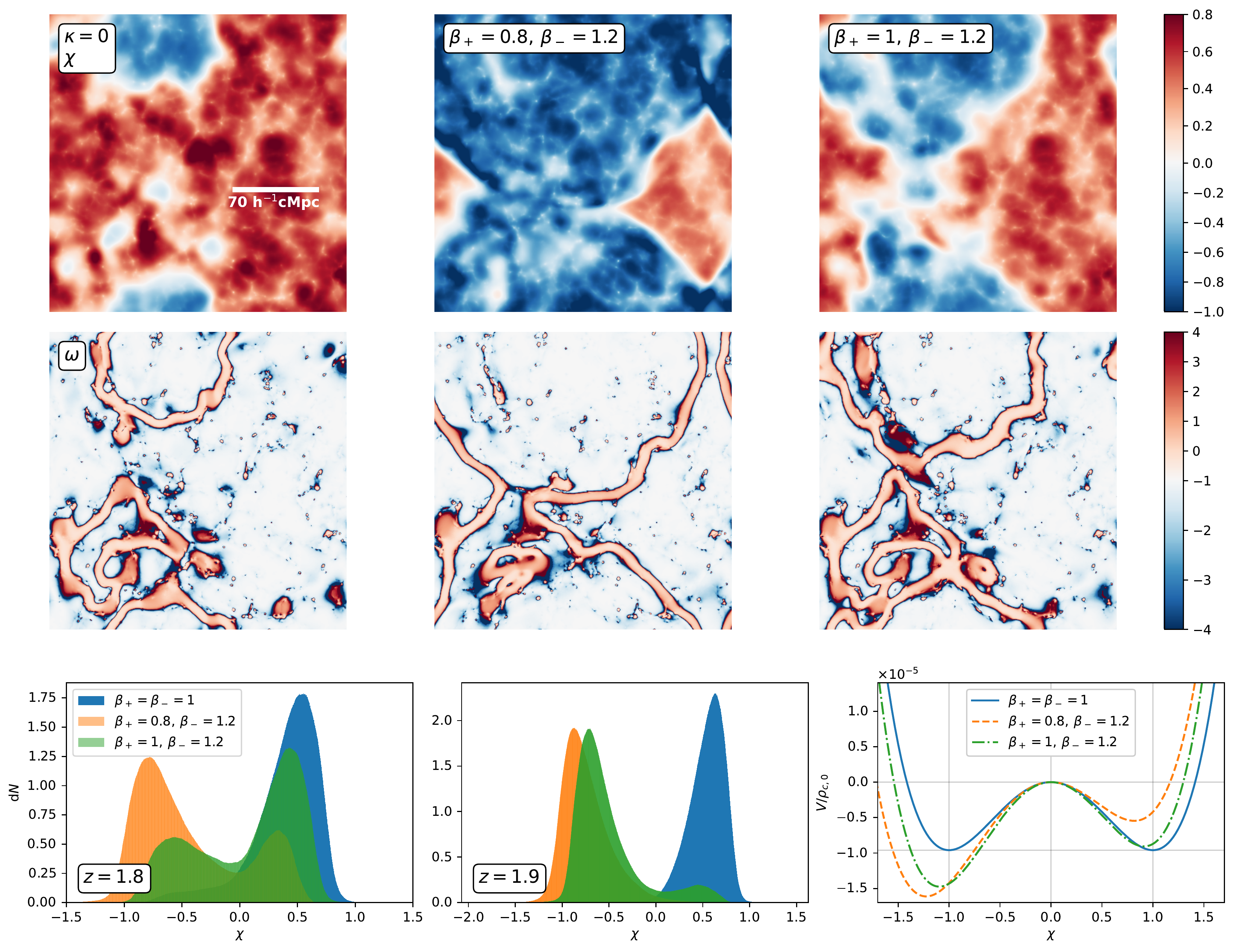}
    \caption{Comparison between asymmetron and symmetron seen through redshift $z=1.9$ cross-section snapshots of the $(256\,\,\,h^{-1}\text{cMpc})^3$ simulation volume, showing the normalised scalar field $\chi$ and the equation of state parameter $\omega$ row-wise. Model shown here is for the symmetron $(\xi_*,a_*,\beta_*)=(3.3\cdot 10^{-4}, 0.33, 1)$, whereas for the asymmetron everything is the same apart from $(\beta_-,\beta_+) = (1.2,0.8)$ in the middle column, and $(\beta_-,\beta_+)=(1.2,1)$ in the rightmost column. In both instances, we choose $\kappa<0$. Bottom row shows in the two figures to the left the histograms of the normalised (a)symmetron field across the boxes at different redshifts. The right figure shows the different potential wells of the two models, where the $y$-axis has potential energy density normalised with respect to the present time critical energy density of the universe}
    \label{fig:asymmetron_comparison}
\end{figure}
In figure \ref{fig:asymmetron_comparison}, we show the result of our comparison of the symmetron model with parameters $(\xi_*,a_*,\beta_*) = (3.3\cdot 10^{-4},0.33,1)$ and the asymmetron models that instead have the two $\beta'$s $(\beta_-,\beta_+)=(1.2,0.8)$, and $\kappa<0$. Results are shown for a (256 $h^{-1}$cMpc$)^3$ volume box size and resolution of $0.5$ $h^{-1}$cMpc. Note that in these cases the presence of the cubic term seems to skew the 
distribution of the field towards the deeper minimum so that the simulation with $\kappa<0$, prefers the negative minimum. In the case of $\kappa>0$, instead, the positive minimum will dominate, as it coincidentally does for the symmetron simulation with $\kappa=0$. 
The equation of state plot, shows some domains in either simulation, but they seem to dissipate faster in the asymmetron scenario than for the symmetron one; Although the asymmetron scenarios start out more distributed among the two minima at $z=1.9$, by $z=1.8$ they are almost entirely thermalised and equally skewed towards their preferred minimum as the symmetron is. The lowermost and right-side plot of figure \ref{fig:asymmetron_comparison}
shows that the negative potential well has become both deeper and more negative while the positive one has become shallower and less positive for $\kappa<0$, whereas the opposite would be the case for $\kappa>0$.  We show the symmetron scenario in more detail in figure \ref{fig:SSB_cartoon}.
\subsection{\oen{Domain sizes}}
\begin{figure}
    \centering
    \includegraphics[width=\linewidth]{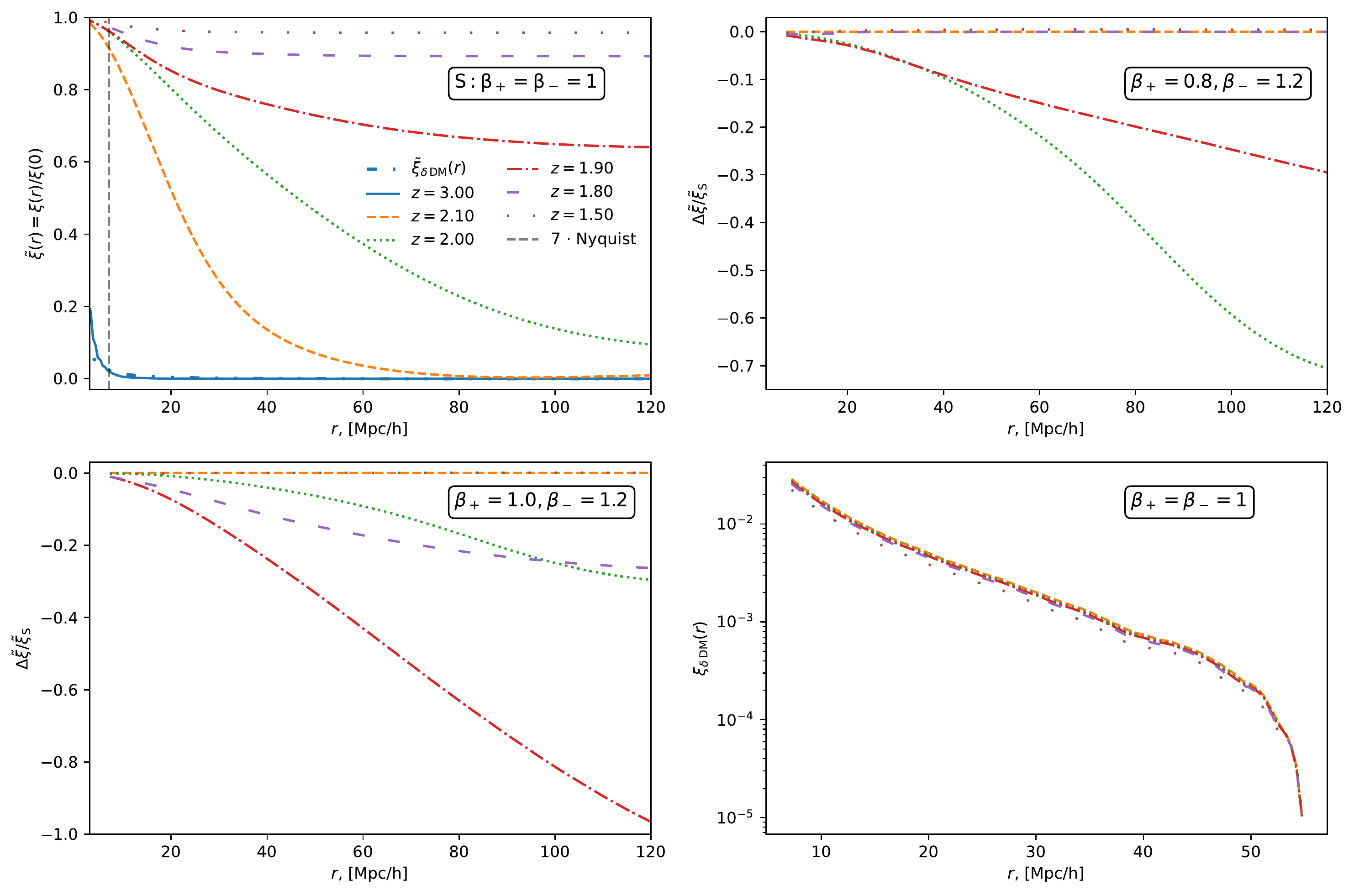}
    \caption{\oen{
    Top left: Monopoles of autocorrelations of the scalar field $\xi=\langle\chi\chi\rangle$, normalised to their values at $r\rightarrow 0$ Mpc/h, to give an indication of the redshift evolution of the domain scales. The corresponding plot for the dark matter autocorrelation monopole is shown as reference, and is redshift independent, since it grows with the growth factor $D_+(a)$. The parameter choice for the symmetron corresponds to $(\xi_*,a_*,\beta_*) = (3.33\cdot 10^{-4},0.33,1)$. The top right and bottom left figures show the relative differences to the symmetron model in the corresponding asymmetron models with parameters $(\beta_+,\beta_-)=(0.8,1.2)$ and $(\beta_+,\beta_-)=(1,1.2)$ respectively. Bottom right: autocorrelation function of dark matter $\xi_{\rm{DM}}=\langle\delta_{\rm{DM}}\delta_{\rm{DM}}\rangle$ as found in the symmetron scenario.
    }}
    \label{fig:domainsize_autocorrelation}
\end{figure}
\noindent
\oen{
In figure \ref{fig:SSB_cartoon} we saw a cartoon indicating the redshift evolution of the domains in the symmetron simulation. By eye, the smaller domain looks to be around $100$ Mpc/h in diameter at $z=1.9$, whereas the larger domain extends across the box. Since we are not resolving a large number of domains, to have accurate quantitative statistics on average domain sizes, we would need to run simulations with larger volumes compared to the domain sizes. The scale of the domains indicate the scale of gravitational correlation for observers attributing the fifth force to gravity, so we are still interested in making quantitative estimation of it for our simulations. We present in figure \ref{fig:domainsize_autocorrelation} the autocorrelation of the scalar field $\xi(r) = \langle\chi\chi\rangle(r)$, normalised to its value at $r \rightarrow 0$. The figure indicates how rapidly the self-correlation drops with larger scales. Before symmetry breaking, there is not much self-correlation for scales much larger than the Compton wavelength, which we expect, and the drop in correlation is similar to that of the dark matter overdensity field. When symmetry breaking happens at $z_* = 2$, the correlation starts growing to larger scales until it completely saturates at $z=1.5$ and there is only one domain left in the simulation volume. A tentative estimate of the domain size, chosen from where $\tilde \xi(r) = \xi(r)/\xi(0)=0.2$, gives the results shown in table \ref{tab:domainscales_estimated}, where we see that the domains quickly extend to larger scales than what is available in the simulation volume.
\begin{table}[!ht]
    \centering
    \begin{tabular}{|c|c|c|c|}\hline 
       $z$ & \multicolumn{3}{|c|}{ Autocorrelation scale, [Mpc/h]}\\\hline
        & $\beta_+=\beta_-=1$ & $\beta_+=0.8,\beta_-=1.2$ & $\beta_+=1,\beta=1.2$ \\\hline
       $2.1$  &  $35$   & $ 35$  & $ 35$ \\
       $2$    &   $ 85$  & $71$ & $ 78 $ \\
       $1.9$ & $>256$  & $>256$ & $87$ \\
       1.8 & $>256$  & $>256$ & $>256$ \\\hline 
    \end{tabular}
    \caption{\oen{Scales where $\tilde \xi=0.2$ as an estimate of the average domain sizes during the time following the spontaneous symmetry breaking. }}
    \label{tab:domainscales_estimated}
\end{table}
}
\newline
\subsection{Background energy transfer}\label{A:energytransfercheck}
The matter stress-energy tensor in the Einstein frame is not conserved, but is sourced by a term which we can evaluate at background level (for $\nu=t$)
\begin{linenomath}
\begin{align}
    \nabla_\mu T^{\mu\,(m)}_{\,\,t} = \dot \rho_m + 3H \rho_m \equiv \mathcal{S},
\end{align}\end{linenomath}
which according to equation \eqref{eq:current} is equal to 
\begin{linenomath}
\begin{align}
\mathcal{S} = \left(\frac{v}{M}\right)^2\langle \rho_m\chi q/A\rangle/a^3,
\end{align}\end{linenomath}
where we are required to perform the average on the product $\rho_m\chi q/A$ right hand side. We can simplify the equation
\begin{linenomath}
\begin{align} \label{eq:rhodmchangeLHS}
      \frac{\diff \left(a^3 \rho_m\right)}{\diff \tau} =  a^4 \mathcal{S} .
\end{align}\end{linenomath}
\begin{figure}
    \centering
    \includegraphics[width=0.6\linewidth]{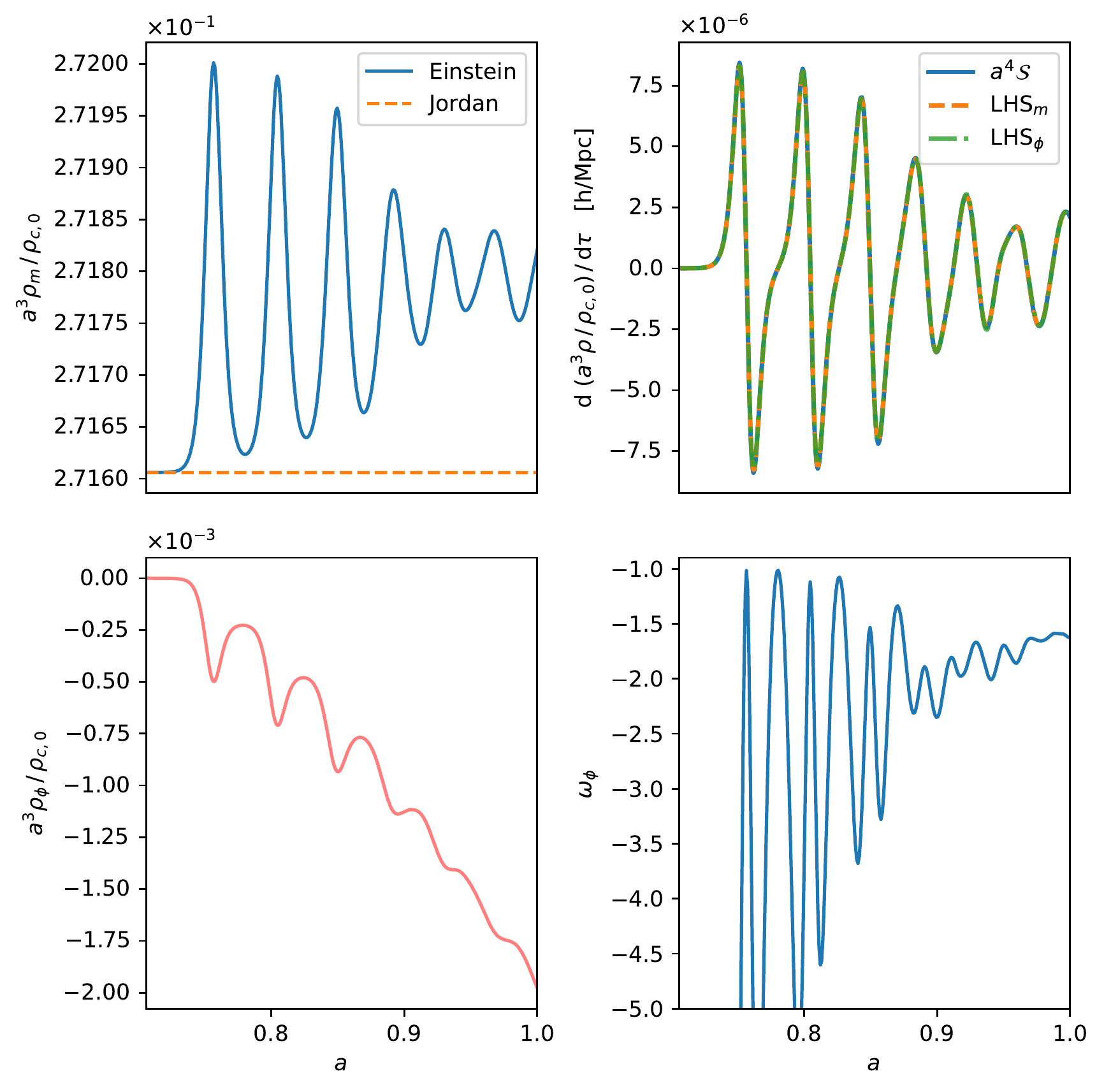}
    \caption{
    The energy transfer between the asymmetron and the matter sector for parameters $(\xi_*,a_*,\beta_*)=(6\cdot 10^{-3},0.33,1)$. In the left panels the energy density parameters $\Omega_i = \rho_i/\rho_{c,0}$, multiplied by $a^3$ for the matter part (top) and the asymmetron (bottom) are plotted. In the top right is the comparison between estimations of the rate of change of these quantities, estimated from the source function $\mathcal{S}$ or the left hand sides of equations \eqref{eq:rhodmchangeLHS} (LHS$_m$) or \eqref{eq:rhoAchangeLHS} (LHS$_\phi$). Bottom right is the equation of state parameter of the asymmetron.}
    \label{fig:energytransfercheck}
\end{figure}\noindent
We consider it as a consistency check to verify that the asymmetron is evolved in such a way that it satisfies equation \eqref{eq:rhodmchangeLHS}. Similarly, the rate of change of energy for the asymmetron stress-energy tensor \eqref{Tmunu_def} can be found
\begin{linenomath}
\begin{align}
    \dot \rho_{\phi} + 3H \rho_{\phi}\left(1+\omega_{\phi}\right)= -\mathcal{S},
\end{align}\end{linenomath}
which, if the equation of state parameter $\omega$ is a constant, we can express differently as
\begin{linenomath}
\begin{align} \label{eq:rhoAchangeLHS}
    -a^{-3\omega}\frac{\diff\left(a^{3(1+\omega)}\rho_{\phi}\right) }{\diff\tau} = a^4 \mathcal{S}.
\end{align}\end{linenomath}
However, $\omega$ varies rapidly for our parameter choice, so instead, we express the equation as
\begin{linenomath}
\begin{align}
    -\frac{\diff \left(a^3 \rho_{\phi}\right)}{\diff\tau} - 3 \mathcal{H} a^3 \rho_{\phi} \omega_{\phi} = a^4\mathcal{S},
\end{align}\end{linenomath}
the left hand side of which we refer to as LHS$_\phi$ in figure \ref{fig:energytransfercheck}. Similarly, we are naming the left hand side of equation \eqref{eq:rhodmchangeLHS} LHS$_{m}$.
We observe a perfect agreement between the different expressions in figure \ref{fig:energytransfercheck} that shows the results from a $(1000$ $h^{-1}$cMpc$)^3$ volume simulation with 7.8 $h^{-1}$cMpc resolution. We comment that the constant Jordan frame energy density plotted left hand side is plotted with respect the Jordan frame scale factor $\tilde a$, or equivalently that it is $A^3\tilde \rho$ that is plotted with respect to the Einstein frame scale factor (see the previous comment around equation \eqref{eq:difficultexplanation}).

Finally we comment that upon inspection of figure \ref{fig:energytransfercheck}, we find $\sim 6.8$ full fluctations from $a\sim0.72$. We calculate the frequency of oscillations
\begin{linenomath}
\begin{align}
    \frac{N}{\Delta\tau_P} = 0.0074
    \text{ h/Mpc},\quad \frac{\mu}{2\pi} = 0.0063\text{ h/Mpc},
\end{align}\end{linenomath}
where $N$ is the number of oscillations during the time interval $\Delta\tau_P$ and $\tau_P$ is used for the particle horizon which relates to the conformal time as $\tau_P = a\tau$. We have written the mass parameter $\mu$ corresponding to $\xi_*= 6\cdot 10^{-3}$ and divided on $2\pi$ for comparison. We note their similarity. Indeed, looking at equation \eqref{eq:lineartest_omegadef}, we see that in the late time when $\rho_m\ll \mu^2 M^2$ and when $\mathcal{H}\ll a\mu$, we expect the oscillation of the linear solution (for large scales $k\to 0$) to have the frequency $a \mu/2\pi$ with respect to conformal time.
We take these findings to pass as consistency checks for the code implementation.

\section{Discussions and conclusion}

\subsection{Choice of cubic term}
For the sake of introducing an asymmetricity in the theory, we could have considered also linear terms, quintic- ($\phi^5$) or higher order terms in the potential, or odd terms in the conformal coupling instead. The reason for having it in the potential instead of the conformal coupling is because in the latter instance, the term would come multiplied with the trace of the stress-energy tensor, and thus be more active in regions of high density. Since the second derivative of any odd term at origin $ V^{(\kappa)}_{\phi\phi} |_{\phi=0}=0$, they do not affect the symmetry breaking criterion. We want to have the effect of asymmetricity expressed in the symmetry broken regions where the asymmetron becomes active, and have therefore placed the odd term in the potential, as in \cite{perivolaropoulos_gravitational_2022}. A linear term (i.e. a tadpole) would be possible, but could not add to the dynamics of the field since one cannot draw up 1 part irreducible Feynman graphs with more than one particle. It can add to the expectation value of the field however, so it might be interesting for a future study. Finally, pentic and higher-order odd terms in the potential would be troublesome in general, since they would become dominant at large enough field-values and make the potential unbound to one of the directions. Having them would necessitate an still higher order even term to make the potential again bounded. Larger terms should be small in practice within the box, since the parameters that will be considered always have vacuum expectation values close to the origin. In general, having an odd term explicitly breaks the parity symmetry of the field and makes it more vulnerable to quantum corrections, as discussed in \cite{hinterbichler_symmetron_2010}. There is therefore a fine tuning problem. However, we view it here agnostically as an effective theory of some more fundamental theory that would have symmetries to guard it against large quantum corrections, which was already the view presented in \cite{hinterbichler_symmetron_2010} since the matter-coupling is non-renormalisable.

\subsection{Validation and prospects}
We see an overall behaviour of the code that is consistent with expectations when applied the idealised scenario in appendix \ref{A:analytic_comparison}. Furthermore, we produce similar results to the established quasi-static $N$-body solver ISIS when running the two with similar initial conditions, as seen in figures \ref{fig:powerspectrum_comparison_isis}, \ref{fig:histograms_comparison_isis} and \ref{fig:slice_comparison_isis}. The qualitative behaviour of the dynamics as shown in figure \ref{fig:SSB_cartoon} seems consistent with what has been presented in \cite{llinares_releasing_2013,llinares_domain_2014}, and the energy conservation equations in subsection \ref{A:energytransfercheck} is demonstrated to be satisfied.

Overall, convergence has been tested with respect to the temporal and spatial resolution, although we are limited by computational resources constraints for finer grids than what is shown in figures \ref{fig:SSB_cartoon} and \ref{fig:asymmetron_comparison}, where we used $(512)^3$ grids for a $(256\,\,\,h^{-1}\text{cMpc})^3$ volume box, meaning that we have a $\sim 0.5\,\,\,h^{-1}$cMpc resolution. Perhaps we need higher resolution still to fully resolve the non-linear structure of the domain walls and their steep gradients. It is possible that their instability is owing to a resolution effect, as commented in \cite{llinares_domain_2014}. The computational challenge and need for resolving a large range of scales for studying domain wall networks is emphasised in \cite{llinares_domain_2014}, where they predict the average distance between domain walls for different redshifts of symmetry breaking \oen{using causality (see chapter 6.4-6.5 of \cite{vachaspati_kinks_2007})}; For $z_*\sim 1$ the average distance should be $\sim 800 $ Mpc/h, and the constraints mentioned in \cite{hinterbichler_symmetron_2010} imply the need for small Compton wavelengths $\sim 1$ Mpc/h -- this requires us to resolve a large range of scales. It might therefore be interesting to compare idealised field configurations in asevolution with the same setup in the SELCIE code \citep{briddon_selcie_2021} that makes use of a finite elements scheme to resolve the steep gradients that are expected to form. This would help to verify that we are sufficiently resolving the domain walls when using a fixed grid separation of half of the Compton wavelength. Producing finer resolution simulations (that contain domain walls) than the ones that we have presented here would be a more direct way to establish this. Stability of the domain wall network and its dynamics is something we wish to study more closely in a later paper, through simulations with larger box sizes and with improved resolutions.

For the relativistic effects, we demonstrated that they were small on the dynamics of the scalar field, for the choice of parameters and simulation volume we made for doing the comparison. The relativistic  effects are more important on large scales, so to study the relativistic corrections one needs cosmological scale simulations \citep{adamek_general_2016}. On another note, it would be interesting to study the effect of the field on the cosmological observables through the past lightcone of an observer in the simulation, by use of which we might find some signature from weak gravitational lensing, integrated Sachs-Wolfe, non-linear Rees-Sciama, gravitational redshift and Shapiro time delay  effects.

Regarding the background part, we seem to be able to find a consistent behaviour and a significant impact of including dynamics for some parameter choices, which would cause the field to halt at origin for some time before falling into its minima in a dampened oscillation. The quasi-static assumption that the transition happens at much smaller than cosmological timescales so that the field effectively always sits in its minima, does not seem to hold for large Compton wavelengths, and can lead to a bump in the Hubble parameter as seen in figure \ref{fig:background_backreaction}, which could alleviate the $\sigma_8$ tension by introducing a period of enhanced expansion slowing down structure formation. It can furthermore be seen in figure \ref{fig:background_backreaction} that the matter energy density parameter is smaller than it is in the $\Lambda$CDM scenario during this transition. The scenario corresponding to the parameter choice of figure \ref{fig:background_backreaction} could however be limited by the argument that $L_C<1$ Mpc/h in order to satisfy local solar system tests as presented in \cite{hinterbichler_symmetron_2010}, although this should be considered more carefully for the exact parameter choice, since they used $a_*\sim 1$, and confronted with new data. The effect on the Hubble rate of  the box-averaged energy density of the dynamic scenario was shown to agree well with that of the dynamically integrated background equation in \eqref{eq:tmp1}, and we therefore see negligible importance of back-reaction in this instance, but there seems to be a larger dampening in the box-averaged field in the bottom right subfigure of figure \ref{fig:background_backreaction}. Understanding whether this small effect truly is owing to some back-reaction of the field or if it is relating to for example the box size, may be interesting to consider more closely in the future. Indeed, although the results shows convergence in spatial and temporal resolution, the agreement between the dynamic background field solution and the averaged dynamic field seems to improve for smaller box sizes, agreeing perfectly for ($200$ $h^{-1}$cMpc)$^3$ volume boxes, keeping the same resolution. Although the background analysis was done with a $\Delta A \sim 10^{-2}$ that thus does not satisfy the criterion in equation \eqref{eq:symmetronGeodesicConstraint}, we expect the imprecision to only affect the part which is performed in the Jordan frame, that only affects clustering. To be sure, we attempted turning the fifth force off and saw no effect on the background analysis results and verified the simulation's consistent behaviour with respect to the test discussed in subsection \ref{A:energytransfercheck}. An improvement of the Jordan frame implementations to be valid for larger $\Delta A$ would be an interesting extension for the future. The physical motivation for such a consideration would however need to rely on the constraints of among other \cite{hinterbichler_symmetron_2010} being dealt with by some mechanism.

\subsection{Conclusion}
We presented an implementation of the (a)symmetron model made on top of the gevolution $N$-body code and validated it with respect to the ISIS code, that solves the symmetron by use of the quasi-static approximation. We have found importance of inclusion of dynamics of the field on the background for large choices of Compton wavelengths. There will possibly also be important effects on clustering, although we did not see a great effect for our code comparison parameter choice where we were using a somewhat small Compton wavelength $L_C=1$ Mpc/h and a later redshift $z=2$ symmetry breaking. We noticed complex dynamics in the box during symmetry breaking, and found that although the cubic term can help to skew the field configuration towards having more domain walls, our results indicate that the domain walls become less stable. A natural next step will be to simultaneously include all relevant effects for both large and small Compton wavelength fields, to reevaluate previous constraints on the symmetron scenario, and to consider whether the asymmetron is able to perform better. The additional phenomenology introduced by including dynamics might also allow us to make stronger constraints by introducing new observational signatures that has not been considered in the past, which we would like to explore further. In particular, studying domain walls in large simulations that are resolving them completely, will answer previous questions on whether they have been sufficiently resolved in the past, and help us to understand how they affect cosmological observables better.

In summary, we have presented the first code to solve the dynamic symmetron with the addition of relativistic degrees of freedom, and the first implementation of the asymmetron that includes a cubic term in the potential. Furthermore, this is to the authors knowledge the first time the background of the (a)symmetron has been solved with back-reaction by averaging of the energy of the field across the box, and we have opened a new avenue of exploration of the non-linear and dynamic behaviour of the model by use of a consistent implementation of the background, sourcing of the Poisson equation and a practical methodology to identify and manipulate the formation of domain walls.


\section*{Acknowledgements}
We would like to thank Hans Winther, Alessandro Casalino, Leandros Perivolaropoulos, Foteini Skara, Claudio Llinares, Mikolaj Szydlarski, Jose Béltran Jiménez and Salome Mtchedlidze for discussions. We thank the Research Council of Norway for their support.  The simulations were performed on resources provided by 
UNINETT Sigma2 -- the National Infrastructure for High Performance Computing and 
Data Storage in Norway. This
work is supported by a grant from the Swiss National Supercomputing Centre
(CSCS) under project ID s1051. Mona Jalilvand acknowledges support through a McGill University postdoctoral fellowship. 

----------------

\bibliography{zPapers,zPapersExtra}

\begin{appendix}

\section{Mapping between Lagrangian and phenomenological parameters}
\label{A:mapping}

In this appendix we discuss the mapping between the parameters $(\mu, M,\lambda,\kappa)\leftrightarrow (\xi_*, a_*,\beta_+,\beta_-)$. One can find the mapping $M\leftrightarrow a_*$ from equation \eqref{eq:SSBpotential}. At the background level $\rho_m = \Omega_m \rho_{c,0}/a^{3}$, so
\begin{linenomath}
\begin{equation}
    \partial^2_\phi V\,\big|_{\phi=0}=0
    \implies a_*^3 = \frac{\rho_{c,0}\Omega_m}{\mu^2 M^2}.
\end{equation}\end{linenomath}
We see therefore that in order to obtain the Lagrangian parameter $M$, we first need to express $\mu$. The mass of the field sets the strength of its lowest order self-interaction, from which we can derive the Yukawa potential
\begin{linenomath}
\begin{equation}
    V = -\frac{g}{4\pi r}e^{-\sqrt{2}\mu r},
\end{equation}\end{linenomath}
where $r$ is the distance between two sources that are coupled to the field.
The potential has a cutoff at the scale $r\sim \frac{1}{\sqrt{2}\mu}$, from which we can define the field's Compton scale $L_c = \frac{1}{\sqrt{2}\mu} $; It can be viewed as a characteristic scale for the field's self-interaction. Defining the parameter $\xi_* = H_0 L_c$, we obtain
\begin{linenomath}
\begin{align}
    \xi_* = \frac{H_0}{\sqrt{2}\mu}. 
\end{align}\end{linenomath}
 The Compton scale is easily recovered using
\begin{linenomath}
\begin{equation}
    L_c = \xi_* \cdot 2998 \text{ Mpc/h}.
\end{equation}\end{linenomath}
Finally, we find $\beta_+,\,\beta_-$, via equation \eqref{eq:betadef}
\begin{linenomath}
\begin{equation}
    \beta_\pm \equiv \frac{\left|v_\pm\right| M_{\text{pl}}}{M^2} = \frac{M_{\text{pl}}}{M^2}\left|\frac{\kappa \pm \sqrt{\kappa^2+4\lambda \mu^2}}{2\lambda}\right|,
\end{equation}\end{linenomath}
where $v_\pm$ is found using \eqref{eq:avev}. The values of $v_\pm$ depend on both the $\lambda$ and $\kappa$ parameters non-linearly. As a result, we might need to solve them numerically using e.g. Newton's method.
However, we can solve for $\lambda$ and $\kappa$ analytically by specifying $\beta_+>\beta_-$, without loss of generality for the model, so that we can uniquely determine $\beta_\pm$ from their average $\bar \beta$ and their difference $\Delta \beta$. There is still a complication since we have to deal with absolute values, but we can remove them by use of knowledge that $\kappa^2<\kappa^2+4\lambda\mu^2$ so that (loosely speaking) $|\beta_-| = -\beta_-$ and $|\beta_+| = \beta_+$ (where right hand sides are supposed to be understood without absolute sign). Now we can find easily the asymmetric part $\beta_+-\beta_-$ and solve for $\kappa$ 
\begin{linenomath}
\begin{align}
    \kappa = \frac{\lambda M^2\Delta\beta}{ M_{\text{pl}} }.
\end{align}\end{linenomath}
We need one more step of algebra. We calculate $\bar \beta=\left(\beta_++\beta_-\right)/2$ 
\begin{linenomath}
\begin{equation}\label{eq:tmptmp1}
    \bar\beta^2 = \left(\frac{M_{\text{pl} }}{M^2}\right)^2\left(\frac{\mu^2}{\lambda} + \frac{M^4\left(\Delta\beta\right)^2}{4 M_{\text{pl}}^2}
    \right),
\end{equation}\end{linenomath}
which results in
\begin{linenomath}
\begin{equation}
    \lambda = 
    \left(\frac{M_{\text{pl}}}{M^2}\right)^2
    \frac{\mu^2}{\bar\beta^2 - \left(\Delta\beta/2\right)^2
    },
\end{equation}\end{linenomath}
that diverges when $\beta_-=0$, so that from equation \eqref{eq:tmptmp1} also $\kappa\rightarrow \infty$.
For the limit where $\beta_+ = \beta_-=\beta$, we find consistently
\begin{linenomath}
\begin{align}
    &\Delta\beta = 0,\quad\bar\beta = \beta,\\
    &\kappa = 0,\\
    &\lambda = \left(\frac{\mu M_{\text{pl}}}{M^2\beta}\right)^2,
\end{align}\end{linenomath}
which are the expected values for the symmetron case.

\section{Comparison with analytic solutions
}\label{A:analytic_comparison}
For the sake of testing the solver, we simplify the equations of motion so that they are analytically solvable. We consider the linearised version of equation \eqref{eq:ELeq1}. We are then left with
\begin{linenomath}
\begin{align}
    \dot q = a \nabla^2 \chi - a^3 \mu^2 \chi \left( 
    \eta - 1 
    \right),
\end{align}\end{linenomath}
where we have $\eta=\rho_m/(\mu M)^2$. Considering the equation in Fourier space, we obtain a coupled system of ordinary differential equations
\begin{linenomath}
\begin{align}
    \frac{\diff  q}{\diff \tau } &= -
    a^2\chi \left(
    k^2 + a^2 \mu^2 [\eta -1]\right), \\
    \frac{\diff \chi}{\diff \tau} &= \frac{q}{a^2},
\end{align}\end{linenomath}
which we can solve by imposing the initial conditions at the initial redshift. The equation system is equivalent to
\begin{linenomath}
\begin{align}\label{eq:tmpequationlabel}
    \frac{\diff^2 \chi}{\diff \tau^2} &= -2\mathcal{H}\frac{\diff \chi}{\diff \tau} 
    - k^2 \chi - a^2 \mu^2 \left(
    \eta -1 
    \right)\chi,
\end{align}\end{linenomath}
which we solve in the following, for the case of an initial Gaussian perturbation. The general solution would be
\begin{linenomath}
\begin{align}
    \chi(\tau) = \mathcal{A}_1 e^{
    -\alpha\tau 
    +i\omega}
    + \mathcal{A}_2 e^{
    -\alpha\tau 
    -i\omega}, \label{generalSolution}
\end{align}\end{linenomath}
where $\alpha,\omega\in\mathbb{R}$ are defined below. We can decompose the solution into a damped sine and cosine. $\mathcal{A}_i$ here are complex so they contain an initial phase. Note that the above solution assumes constant $\alpha,\omega$ which is not the case since the scale factor and $\mathcal{H}$ are variable. The solution is instead a valid approximation for smaller than cosmological timescales. The analytic solution comparison will then only be a good test for non-cosmological evolution or small redshift intervals.
\subsection{Initial Gaussian}
From equation \eqref{generalSolution} we obtain the solution in Fourier space (having made a choice of initial conditions so that the cosine has zero initial phase)
\begin{linenomath}
\begin{align}
    \chi &= \mathcal{A}\cos(\omega [\tau-\tau_0])e^{-\alpha (\tau-\tau_0)}, \\ \label{eq:lineartest_omegadef}
    \omega^2 &= k^2 + a^2\mu^2 ( \eta - 1 ) +  \mathcal{H}^2,\\ \label{eq:lineartest_alphadef}
    \alpha  &= \mathcal{H}.
\end{align}\end{linenomath}
$\mathcal{A}$ can be a function of $k$, so that we are free to put general initial conditions on the initial slice.
Since asevolution is in real space and we want we simple initial conditions in both $k$-space and real space, we make use of an initial Gaussian in the centre of the box at $z=z_{\text{ini}}$. The Fourier of a Gaussian is still a Gaussian, so we have
\begin{linenomath}
\begin{align}
    \chi(r,z=z_{\rm{ini}}) &= \mathcal{B} e^{-b r^2},\\
    q(r,z=z_{\rm{ini}})  &=  -\alpha\,a^2\, \chi, \\
    \chi(k,z=z_{\rm{ini}}) &= \mathcal{B} \left(\frac{\pi}{b}\right)^{3/2} e^{-\frac{k^2}{4b}},\\
    q(k,z=z_{\rm{ini}}) &= -\alpha\,a^2\, \chi,
\end{align}\end{linenomath}
where $\mathcal{B},b\in \mathbb{R}$ are constants. We obtain the analytic expression for the $\chi$ power spectrum
\begin{linenomath}
\begin{align}
    P(k) = \frac{k^3}{2\pi^2 V}\, \chi_k^2 .
\end{align}\end{linenomath}
In order to get an idea of whether we are resolving the dynamics of the partial differential equation, we use the Courant-Friedrichs-Lewy criterion for the timestep
\begin{linenomath}
\begin{align}\label{eq:supertmp}
    C = \frac{\diff \tau_{\phi}}{\diff x} v_g \leq C_{\text{max}}\equiv C_{\phi},
\end{align}\end{linenomath}
where we have defined the scalar field Courant factor $C_{\phi}$.
From equation \eqref{eq:supertmp} it follows that
\begin{linenomath}
\begin{align}
    \diff \tau_{\phi} \leq C_{\phi} \frac{ \diff x}{v},
\end{align}\end{linenomath}
where we choose $v$ to be the largest of the group velocity 
\begin{linenomath}
\begin{align}
    v_g = \frac{\diff \omega}{\diff k}
    = \frac{k}{\sqrt{k^2 + a^2 \mu^2 (\eta -1) + \mathcal{H}^2}},
\end{align}\end{linenomath}
and the phase velocity\footnote{We note that this is always superluminal prior to symmetry breaking. The group velocity can become complex post symmetry breaking; One should then perturb the field around the new vacuum to find the new physical degree of freedom.}
\begin{linenomath}
\begin{align}
    v_p = \frac{\omega}{k} = \sqrt{1 + \frac{a^2\mu^2 (\eta-1)+\alpha^2}{k^2}},
\end{align}\end{linenomath}
which is at maximum for the smallest mode $k=2\pi/L$ in the simulation box, where $L$ is the box size.
We adjust the time step dynamically in the code, so that the number of internal time loops of the field equation solver $N=\diff\tau/\diff\tau_{\phi}$ with regards to the time step $\diff\tau$ in the external time loop that solves the rest of the fields is chosen every external loop as
\begin{linenomath}
\begin{align}
    N = \left\lceil \text{max}\left( 
    \frac{\diff \tau}{C_{\phi} \diff x}\, v_{\text{max}}
    ,\, N_{\text{choice}},\,1
    \right)\right\rceil.
\end{align}\end{linenomath}
This will be helpful to get an intuition of what are the physical timescales we need to resolve. However, we also looked for convergence of the system with finer time resolutions.

\begin{figure}[!ht]
    \centering
    \includegraphics[width=0.8\linewidth]{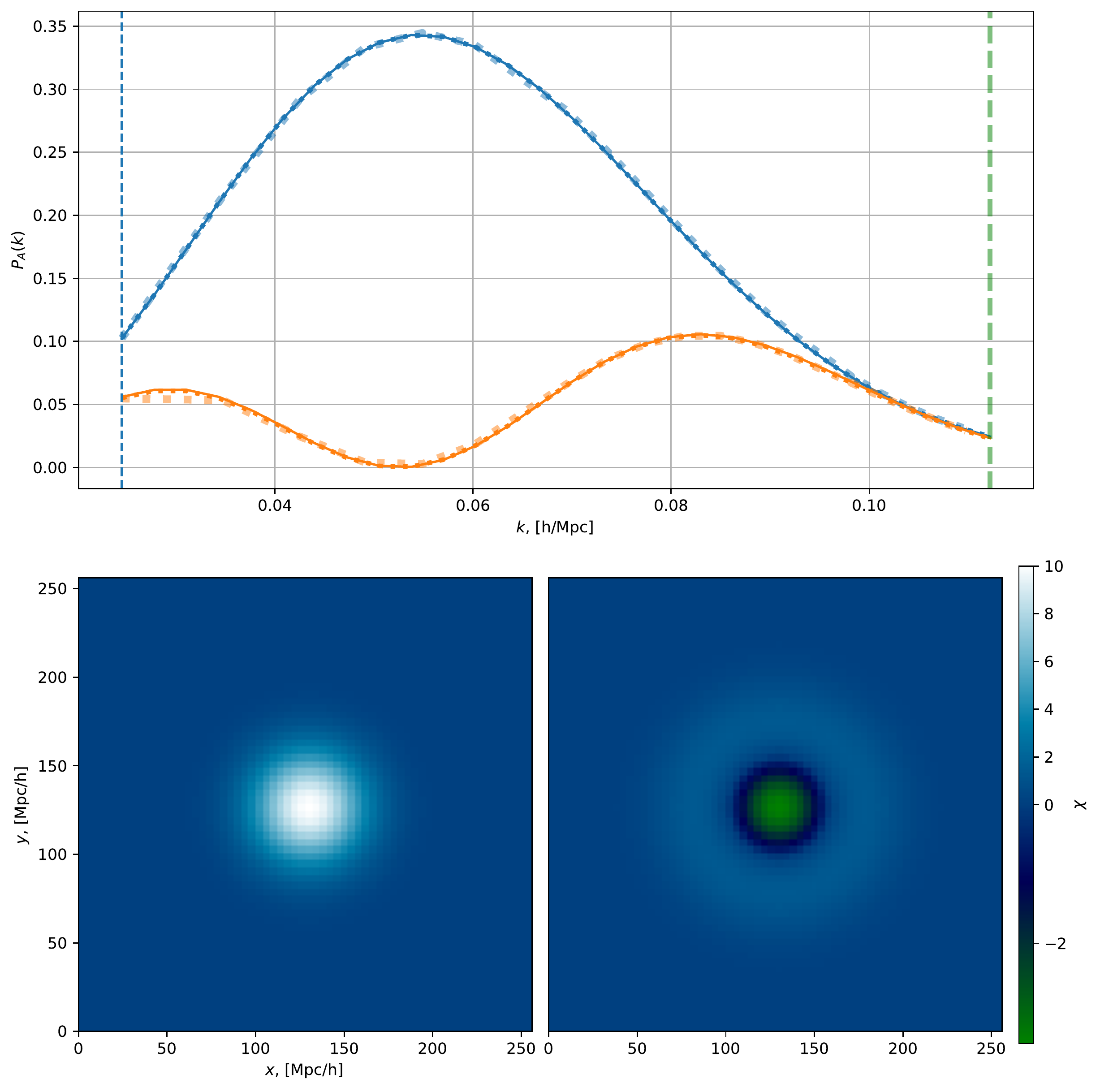}
    \caption{ Top: Comparison of dimensionless power spectra from the Fourier space Euler-Cromer solver (straight line), analytic linear solution (dots) and the asevolution real space leapfrog solver (squares). The scale of the box is indicated with a vertical dashed line on the left side, whereas the scale of one seventh of the Nyquist is indicated with the vertical line on the right side. Bottom: Indicated field configuration, seen from a slice through the centre of the simulation volume, at the start of the simulation (left) and at the end (right).
    }
    \label{fig:analyticPowerSpec}
\end{figure}

We plot the result for the coupled equation system solved in $k$-space together with the analytic result and the result of the simulation in figure \ref{fig:analyticPowerSpec}. The simulation is performed without the scalar field contributing to the stress energy tensor and without fifth forces, and instead of evaluating the local matter density in $\eta$, the background one is used. All relevant parameters used in the simulation are shown in table \ref{tab:GaussiantestParameters}, where $C_f$ is the Courant factor used in the gevolution loop. We note that the choice of a very small $C_f$ is because it sets the time resolution of the output of the power spectrum, and that the scalar field evolves so rapidly that the resolution on the output needs to be small in order to recover the correct scalar power spectrum. Another way to resolve this is to set the output for the scalar field in the internal loop instead, but since the simulation is relatively cheap, we have chosen to put the $C_f$ small.
\begin{table}[!ht]
    \centering
    \begin{tabular}{|c|c|c|c|c|}\hline
        $z_{\text{init}}$ 
        & $C$ & $b$ & $C_f$ & $C_{\phi}$ \\\hline
        0.01& 10 & $10^{-3}$ & $0.48$
        &1   \\\hline
        \multicolumn{5}{c}{}
        \\\hline
        box size & $N_{\rm{grid}}$ & $N_{\rm{pcls}}$ &
        $\xi_*$ & $N_{\rm{choice}}$ \\\hline
        256 $h^{-1}$cMpc & $64^3$ & $64^3$ &
        $0.01$ & 1 \\  \hline      \multicolumn{5}{c}{}
    \end{tabular}
    \caption{Parameters chosen for the Gaussian test case run.}
    \label{tab:GaussiantestParameters}
\end{table}

\section{Units}

For convenience of users and for a better understanding of the dimensions of the parameters, we include a short appendix on the units used in the code and the conversion into them.

The gevolution code uses a unit system where the speed of light $c=1$ and $4\pi G_N=\frac{3H_0^2}{2\rho_{c,0}}\equiv \text{fourpiG} = \frac{3}{2}\left(H_0 L\right)^2= \frac{3}{2}\frac{10^{10}L^2}{c^{2}}$, where $L$ is the box size\footnote{Since we are restoring the speed of light $c\neq1$ in this appendix only, be careful to not confuse it with the c in the comoving units $h^{-1}$cMpc, that does not relate to the speed of light.} in $h^{-1}$cMpc and the speed of light $c$ is written in m/s. The energy density\footnote{Internally in the code there is also a normalisation of the different components of the stress-energy by multiplications of the scale factor.} has been normalised so that $\rho_{i,0} = \Omega_{i,0}$. In other words $\rho_{c,0} = \frac{3H_0^2}{8\pi G_N}=1$ and therefore $H_0 = \sqrt{\frac{2 \cdot \text{fourpiG}}{3}}$. The reason for this choice of fourpiG is because this is exactly what gives a critical density equal to one in gevolution units, which one can see from the conversion factor in table \ref{tab:unit_table}, which is equal to $1/\rho_c$ in Mpc$^2$/h$^2$ (note that this is true when converting from Planck units where $8\pi G_N=1$). Finally, position has been normalised with the box size so that it takes values between 0 and 1, and likewise the conformal time $\tau$ has been normalised by the box size.

To put our field parameters into this unit system, we need to identify their dimensionality. From power counting, knowing that the Lagrangian density is an energy density $\mathcal{L}\sim$J/m$^3$ (in SI), we find the parameters' dimensionalities in some different unit systems shown in table \ref{tab:unit_table}.
\begin{table}[!ht]
    \centering
    \begin{tabular}{|c|c|c|c|c|}\hline
         symbol & SI & $\hbar=1^{(\rm{m})}$ & $G_N=1^{(\rm{m})}$ & conversion  \\ \hline
         $\phi$ & $\sqrt{\text{J/m}}$ & 1/m & 1 & $1/\sqrt{2\cdot \text{fourpiG}}$ \\
         $\mu$ & 1/m & 1/m & 1/m & $L$\\
         $M$ & $\sqrt{\text{J/m}}$ & 1/m & 1 & $1/\sqrt{2\cdot \text{fourpiG}}$ \\
         $\lambda$ & $\frac{1}{\text{J m}}$ & 1 & 1/m$^2$ & $2 L^2\cdot \text{fourpiG}$ \\
         $\kappa$ & $1/\sqrt{\text{J m}^3}$ & 1/m & 1/m$^2$ & $L^2\sqrt{2\cdot \text{fourpiG}}$ \\
         $\rho_c$ & J/m$^3$ & 1/m$^4$ & 1/m$^2$ & $L^2/\left(2\cdot \text{fourpiG}\right)$
         \\ \hline
         \multicolumn{5}{c}{}
    \end{tabular}
    \caption{Dimensionality of the asymmetron parameters and other quantities in three different unit systems (all with $c=1$). The non-SI unit systems are both defined in terms of distance as their only unit. Last column shows the conversion factor that needs to be multiplied to get gevolution units from Planck units. }
    \label{tab:unit_table}
\end{table}
Since in gevolution we have not defined $G_N=1$ but rather $4 \pi G_N =$fourpiG, which is defined above, then we need to be careful to include the appropriate factors as we convert equations from the SI units to the gevolution unit system. In order to convert the SI dimension of, take for example $\phi$ into $\sim 1$ in the gravitational units, we need to multiply by the gravitational constant. Since we are setting it equal to 1, normally we would multiply $\sqrt{\frac{G_N}{1}}$ and then absorb the constant into the field (setting $c=1$ makes the factor $\sqrt{G_N/c^2}$). Now instead, we are setting $4\pi G_N=$fourpiG, so we multiply $\sqrt{\frac{4\pi G_N}{\text{fourpiG}}}$ and absorb $4\pi G_N$ into the field. The corresponding conversion factors for all the parameters can be found in table \ref{tab:unit_table}, where we also have included the normalisation with respect to the box size, $L$.

\end{appendix}
\end{document}